\documentclass{article}

    \usepackage[preprint,nonatbib]{neurips_2022}

\usepackage[symbol]{footmisc}

\usepackage[utf8]{inputenc} 
\usepackage[T1]{fontenc}    
\usepackage{url}            
\usepackage{booktabs}       
\usepackage{amsfonts}       
\usepackage{nicefrac}       
\usepackage{microtype}      
\usepackage{xcolor}         
\usepackage{graphicx}

\newcommand{\simevals}{\texttt{SimEvals}}
\newcommand{\simeval}{\texttt{SimEval}}

\usepackage{color-edits}
\usepackage{algorithm,algorithmic}
\addauthor{nj}{red}
\addauthor{vc}{blue}
\addauthor{gp}{purple}
\addauthor{at}{orange}
\newcommand{\vx}{\mathbf{x}}

\usepackage{amsmath}
\usepackage{amssymb}
\usepackage{mathtools}
\usepackage{enumitem}
\usepackage{amsthm}
\usepackage{bbm}
\theoremstyle{plain}
\newtheorem{theorem}{Theorem}[section]

\theoremstyle{definition}
\newtheorem{definition}[theorem]{Definition}

\PassOptionsToPackage{sort&compress}{natbib}
\usepackage[numbers]{natbib}
\usepackage{hyperref}       

\makeatletter

\title{Use-Case-Grounded Simulations for\\ Explanation Evaluation}

\renewcommand{\thefootnote}{\fnsymbol{footnote}}

\author{%
Valerie Chen\thanks{Correspondence to \texttt{valeriechen@cmu.edu}.} $\:$ Nari Johnson $\,$  Nicholay Topin\footnotemark[2] $\,$ Gregory Plumb\footnotemark[2] $\,$  Ameet Talwalkar \\
 Carnegie Mellon University
}

\begin{document}

\maketitle

\footnotetext[2]{Equal Contribution}

\renewcommand*{\thefootnote}{\arabic{footnote}}

\setcounter{footnote}{0}

\begin{abstract} 

A growing body of research runs human subject evaluations to study whether providing users with explanations of machine learning models can help them with practical real-world use cases.
However, running user studies is challenging and costly, and consequently each study typically only evaluates a limited number of different settings, e.g., studies often only evaluate a few arbitrarily selected explanation methods.  To address these challenges and aid user study design, we introduce Use-Case-Grounded \textbf{Sim}ulated \textbf{Eval}uations (\simevals).
\simevals{} involve training algorithmic agents that take as input the information content (such as model explanations) that would be presented to each participant in a human subject study, to predict answers to the
use case of interest.
The algorithmic agent's test set accuracy provides a measure of the predictiveness of the information content for the downstream use case.  
We run a comprehensive evaluation on three real-world use cases (forward simulation, model debugging, and counterfactual reasoning) to demonstrate that \simevals{} \emph{can} effectively identify which explanation methods will help humans for each use case.  These results provide evidence that \simevals{} can be used to efficiently screen an important set of user study design decisions,
e.g. selecting which explanations should be presented to the user, 
before running a potentially costly user study.

\end{abstract}

\section{Introduction}\label{sec:intro}
\vspace{-0.15cm}

The field of interpretable machine learning has proposed a large and diverse number of techniques to explain model behavior (e.g.,~\citep{ribeiro2016should,lakkaraju2016interpretable, ribeiro2018anchors, plumb2018model, lundberg2017unified}). A growing body of work studies which of these techniques can help humans with practical real-world use cases~\citep{debuggingtests, interpretinginterp, hase_evaluating_2020, pedropaper, expo,poursabzi2018manipulating, chandrasekaran2018do, hase_evaluating_2020}. 
Because it is difficult to anticipate exactly which explanations may help humans with a particular use case~\citep{whitepaper, davis2020measure}, the gold standard to evaluate an explanation's usefulness to humans is to run a \emph{human subject study}~\citep{doshi2017towards}.  These studies ask users a series of questions about the use case and provide the users with relevant information (such as model explanations) to answer each question.


\begin{figure*}[h!]
\centering
  \includegraphics[scale=0.8]{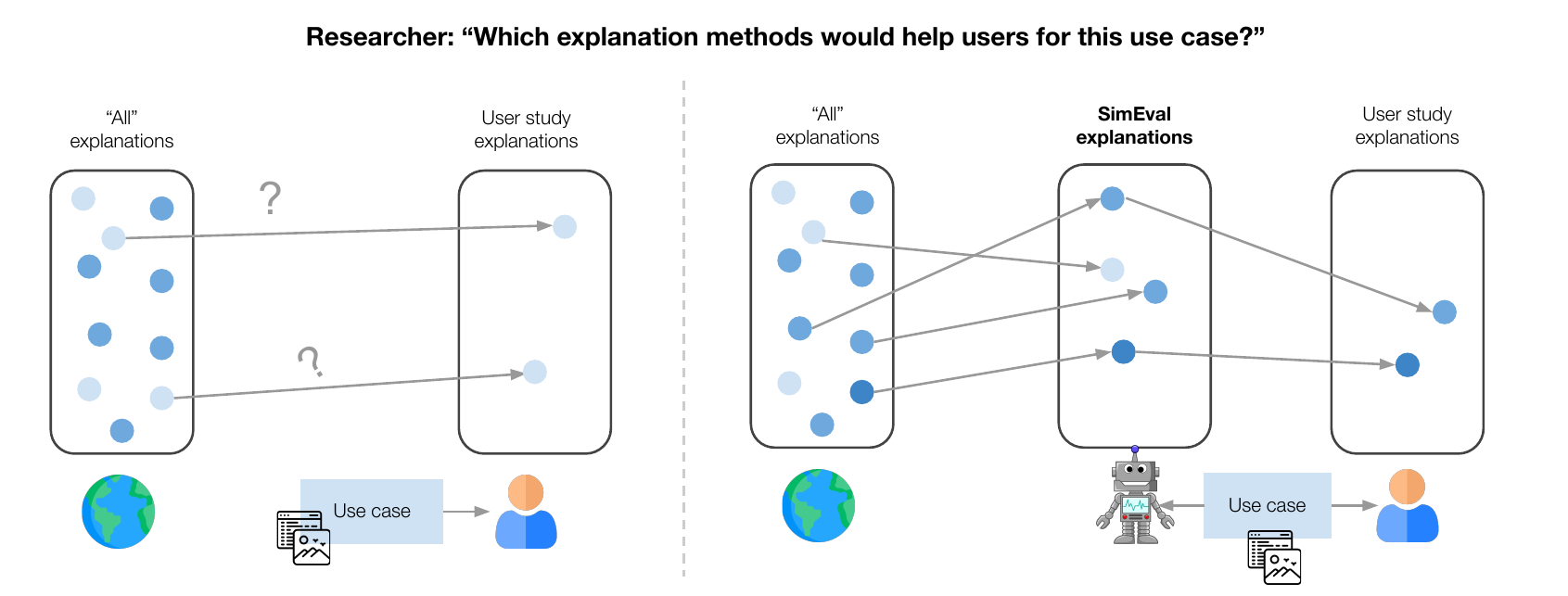}
  \caption{An overview of how \simevals{} can help a researcher with an important user study design decision:  selecting which explanation methods to evaluate given their specific \emph{use case}.
  (Left) Prior to our work, existing user studies often only evaluate a small number of explanation methods due to resource constraints.  When selecting candidate explanations to evaluate, researchers often simply choose the most popular or well known explanation methods with little justification about \emph{why} each explanation may be helpful for the downstream use case.  
  (Right) We propose using \simevals{}, which are use-case-grounded, algorithmic evaluations, to efficiently screen explanations \emph{before} running a user study. In this example, the researcher runs a \simeval{} on each of the four candidate explanation methods and then uses the results of the \simevals{} to select two promising explanation methods where the algorithmic agent has high accuracy for their human subject study.}
  \label{fig:problem}
  \vspace{-0.35cm}
\end{figure*}

Selecting the relevant information, which we call \emph{information content}, to provide each user in the human subject study comprises an important set of design decisions~\citep{pedropaper, bansal2020sam}.
The researcher needs to consider the choice of explanation methods, hyperparameters that are used to calculate the model explanations, and whether additional pieces of information should be provided to the user (such as the model’s prediction on the data-point being explained).\footnote{We note that there are several additional user study design decisions beyond \emph{what information} is shown to the user: for example, how the information is presented or visualized, the user study interface design, etc.  We consider these additional design decisions to be out-of-scope for our study for reasons discussed in Appendix \ref{appdx:humanfactors}.}
While a researcher may wish to run several user studies to evaluate many different types of information content, in practice this is often infeasible: user studies are resource intensive and time-consuming, requiring the recruitment and compensation of real users. Due to these constraints, most studies are far from comprehensive in terms of the number of design decisions evaluated.
For example, most user studies typically only evaluate a few explanation methods, with default hyperparameter settings, that are often arbitrarily selected based on their perceived popularity (Figure~\ref{fig:problem}, Left). 

For example, forward simulation is a canonical use case where users are asked ``Can you predict what the model's prediction is for this individual?''~\citep{poursabzi2018manipulating, chandrasekaran2018do, hase_evaluating_2020}. To answer this question, the user is provided with information content, which may include data-points (e.g., the individual's features) and model explanations for those data-points (e.g., a SHAP~\citep{lundberg2017unified} importance score for each feature). The researcher can then measure which set of information content (and consequently which \emph{explanation method}) enables users to achieve the best performance (e.g. the highest accuracy when answering the use case questions).
Beyond forward simulation, other user studies have investigated use cases ranging from model debugging to counterfactual reasoning~\citep{debuggingtests, interpretinginterp, hase_evaluating_2020, pedropaper, expo}.  

In this work, we introduce Use-Case-Grounded \textbf{Sim}ulated \textbf{Eval}uations (\simevals{}): a general framework for conducting use-case-grounded, algorithmic evaluations of the information content present in a user study (Figure~\ref{fig:problem}, Right). Each \simeval{} involves training and evaluating an algorithmic agent, which is an ML model. The agent learns to predict the ground truth label for the use case given the same information that would be presented to a user in a human subject study. The agent's test set accuracy is a measure of whether the information content provided (which critically includes the model explanation) is predictive for the use case.  

By comparing the agent's test set accuracy for different types of information content, the researcher can efficiently identify promising and eliminate unhelpful information content types (e.g. explanation methods as shown in Figure~\ref{fig:problem}) to present to real humans in their final human subject study.  Since \simevals{} are intended to inform, \emph{not} replace, human subject evaluations, we note that our algorithmic agent does not incorporate additional factors such as the effects of cognitive biases~\citep{lage2019evaluation, poursabzi2018manipulating}, that may affect human decision-making (see Appendix \ref{appdx:humanfactors}).  In our experiments, we show that using \simevals{} to measure predictive information \emph{can} indeed identify which information content can help a human with a downstream use case.

In this work, we focus primarily on the design choice of selecting which explanation methods to evaluate. We conduct an evaluation on a diverse set of use cases and explanation methods:  (1) We compare \simevals{} with findings from prior user studies, and (2) we run our own user study motivated by use cases studied in prior work.  In both cases, we find that \simeval{} effectively distinguishes between explanations that are ultimately helpful versus unhelpful to human subjects in the user study.

Our primary contributions in this work are as follows:

1. \textbf{Introducing \simevals{} Framework.} We are the first to propose an automated use-case-grounded evaluation framework to measure the predictive information contained in the information content provided to a user in a human subject study.
We instantiate our framework on three diverse use cases to illustrate important differences in how \simevals{} can be instantiated, showcasing how the \simeval{} framework covers a wide variety of real-world use cases and explanation methods.

2. \textbf{Identifying Promising Explanations for Humans.} We demonstrate that \simevals{} can distinguish between which explanations are promising versus unhelpful for humans on each of the three use cases we consider. 
Specifically, we find that when there is a significant gap between \simeval{} performance for two explanations, we observe a similar significant gap in human performance when using the same explanations.

\vspace{-0.15cm}
\section{Related Work}
\vspace{-0.15cm}

\textbf{Types of User Studies on Explanations:}  Many prior works run user studies to evaluate IML methods, covering a diverse set of goals~\citep{mohseni2020multidisciplinary}.  Specifically, IML user studies can differ in the independent variable that is changed (e.g., the explanation method used \citep{interpretinginterp, hase_evaluating_2020} or the visualization of the explanation \citep{krause2018user}) and the dependent variable(s) being measured (e.g., the user's performance on a use case \citep{interpretinginterp, hase_evaluating_2020, lage2019evaluation, lakkaraju2016interpretable, lakkaraju2019faithful, feng2019can}, time taken to complete the use case \citep{lage2019evaluation}, or trust in the ML model \citep{bucinca2020proxy}). We focus on the set of user studies that measure how different \emph{explanation methods} (independent variable) affect how \emph{accurately} users can complete a use case (dependent variable).

We group existing user studies on explanations that measure the accuracy of users on a use case into two categories. The first category studies the utility of explanations in use cases where the \emph{predictiveness} of the explanation method is already known to researchers. For example, ~\citep{lage2019evaluation, lakkaraju2016interpretable, lakkaraju2019faithful} ask users to answer simple comprehension questions about a decision set explanation, where these explanations contain the information necessary to complete the use case with perfect accuracy. Thus, these studies measure humans' ability to comprehend and interpret the explanations without error. In contrast, the second category includes studies where the predictiveness of the explanation method is \emph{not} known to researchers in advance of their study \citep{krause2018user, hase_evaluating_2020, debuggingtests}. Our framework was developed to aid researchers in this second setting to quickly validate whether or not an explanation method is predictive for their use case. 

\textbf{Algorithmic frameworks for evaluating explanations:} 
One line of evaluation frameworks focuses on automating the ability to measure desirable properties relating to the faithfulness of the explanation to the model \citep{kim2021sanity, adebayo2018sanity, bim}, which does not account for any downstream use cases for the explanation. In contrast, our algorithmic evaluation framework allows researchers to customize and perform a \emph{use case grounded} evaluation of explanation methods.

We were able to identify one-off examples that use algorithmic agents to perform use case grounded evaluations in prior work \citep{expo, rps, koh2017understanding, pruthi2020evaluating}.  All of these examples propose evaluations designed around one specific use case. Our framework is more general, and can be applied to a wide variety of use cases for explanations. Our work is the first to propose a learning-based agent to identify which explanation methods may be predictive for a downstream use case.  We provide a more in-depth discussion of how these examples can be contextualized in our general framework in Appendix~\ref{appdx:existingwork}.



\vspace{-0.15cm}
\section{General Framework}\label{generalframework}
\vspace{-0.15cm}

\begin{figure*}[h!]
\centering
  \includegraphics[scale=0.55]{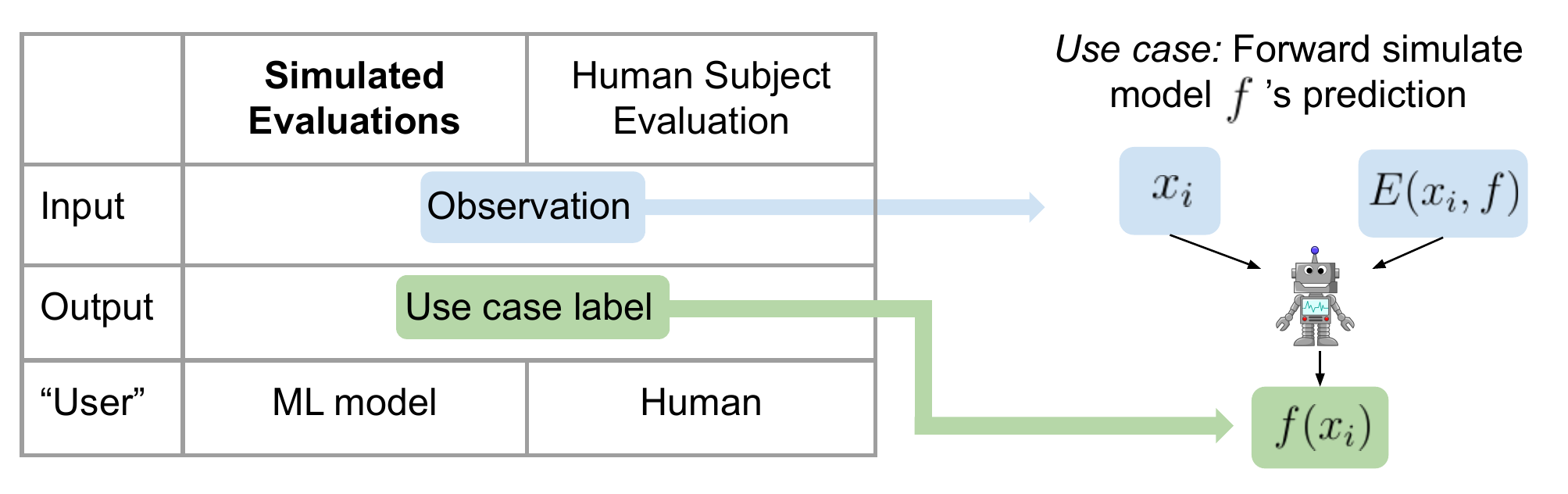}
  \caption{(Left) \simeval{} trains an agent (e.g., ML model) on the same information (observation and use case label) that a user would see in a human subject evaluation.
  (Right) For the forward simulation use case, where the goal is to simulate the ML model's prediction for a given data-point, we demonstrate how to instantiate the observation (which contains the data-point $\vx_i$ and its corresponding explanation $E(\vx_i,f)$) and use case label ($f(\vx_i)$).  This input/output definition is used to generate data to train and evaluate the agent. In Section \ref{experiments} we show that training an algorithmic agent using this framework allows us to corroborate findings about the usefulness of explanations as studied in Hase and Bansal~\citep{hase_evaluating_2020}.}
  \label{fig:simevaloverview}
  \vspace{-0.25cm}
\end{figure*}



We present a general framework to train \simevals{}, which measure the predictiveness of information content choices for a downstream use case. Recall that when designing a user study, the researcher must identify candidate choices of information content (e.g. which explanation methods, hyperparameters, or other additional baseline information) to evaluate. 
Our framework trains a \simeval{} agent for \emph{each} choice of information content that the researcher intends to evaluate for a downstream use case. The researcher can then interpret the test set accuracy of each \simeval{} agent as a measure of the predictiveness of that information content. Before describing how to train each agent, we overview three use-case-specific components  \emph{shared} across all agents/types of information content that the researcher must instantiate:

1. A base dataset ($\mathcal{D} = \{ (\vx,y)\}$), on which the explanation method's utility will be evaluated.\\
2. The base model family, $\mathcal{F}$, which is the family of prediction models. The trained base model $f \in \mathcal{F}$, which is trained on $\mathcal{D}$, is used to generate explanations.\\
3. A function that defines a use case label, which is a ground truth label for the use case. This label corresponds to the correct answer for each question that a user must answer in the user study.

Now, we describe how to train and evaluate each agent model.\footnote{Note the distinction between the \emph{prediction model} being explained (which we call the ``base model'') and the separate \emph{agent model}.} We introduce a three-step general framework: (1) data generation, (2) agent training, and (3) agent evaluation. Since the three steps of our framework closely parallel the steps in a standard ML workflow for model training and evaluation, canonical ML training techniques can also be used to train an algorithmic agent. The primary difference from the standard ML workflow is that in our framework, Step (1) is much more involved to ensure that the data generated as input to the algorithmic agent reflects the information content that would be presented to a human subject in a user study as shown in Figure \ref{fig:simevaloverview}. 




\textbf{Step 1: Data Generation.} The goal of this step is to use the researcher's choice of information content and use-case-specific components to generate a dataset to train and evaluate the agent. The dataset consists of observation and use case label pairs, where the information content defines \emph{what information} is included in each observation . 
We show a concrete example of data generation for the forward simulation use case in Figure~\ref{fig:datagen}. The specific construction of each observation can vary depending on several factors to accommodate a diverse range of use cases and user study designs: 

\begin{figure}[h!]
 \vspace{-0.25cm}
\centering
  \includegraphics[scale=0.55]{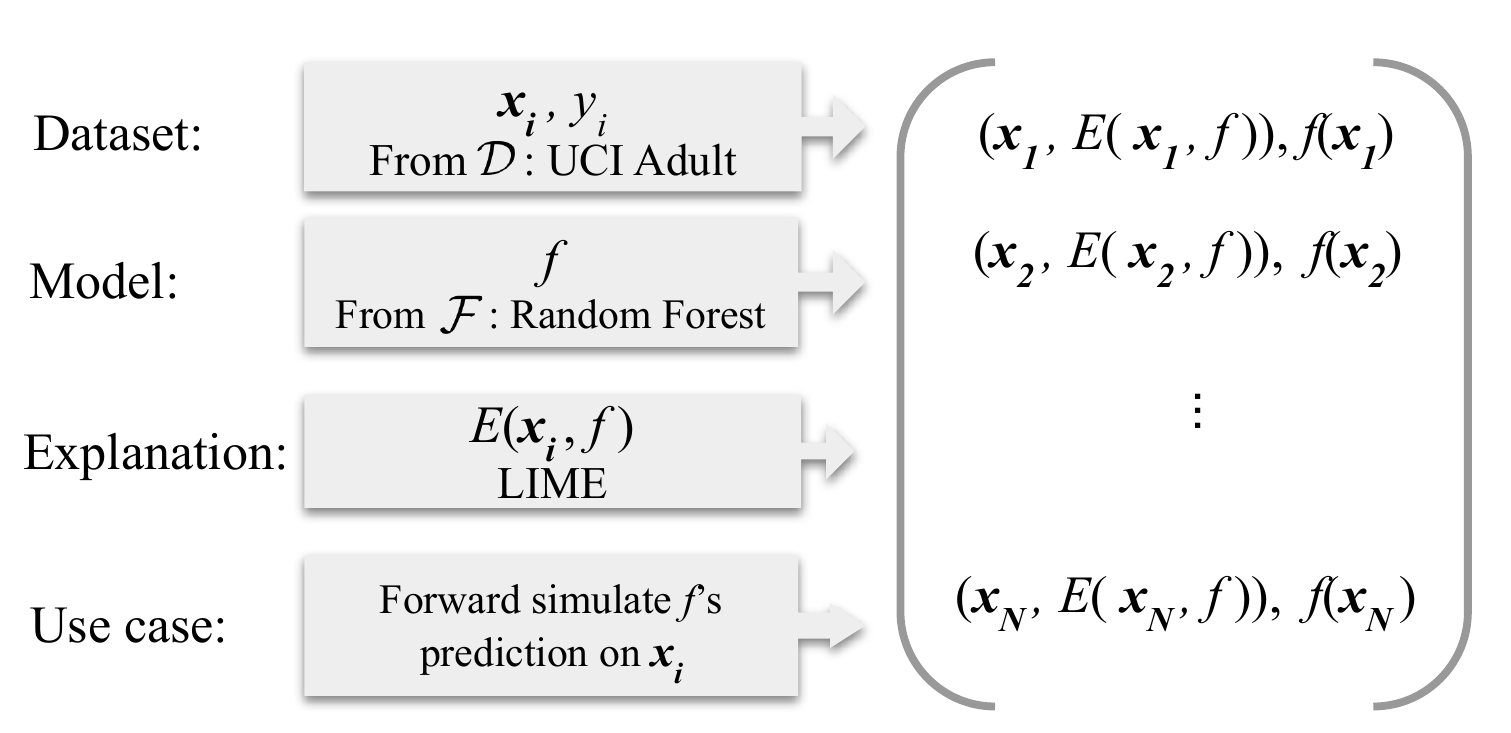}
  \caption{An example of the data generation process for the forward simulation use case from Hase and Bansal~\citep{hase_evaluating_2020}. (Left) For this particular information content type, each observation contains the data-point $\vx_i$ and the model explanation for that datapoint $E(f,\vx_i)$. The three use-case-specific components are the base dataset $\mathcal{D}$, base model family $\mathcal{F}$, and function that defines a use case label. (Right) These components are used to generate a dataset of $N$ observations.}
  \label{fig:datagen}
\end{figure}  

\textit{Local vs. global.}  The \simeval{} framework can be used to evaluate both local and global explanations: A local explanation method $E(f, \vx)$ takes as input prediction model $f$ and a single data-point $\vx \in \mathcal{X}$. A global explanation method $E(f, \mathcal{D})$ takes as the input model $f$ and a dataset $\mathcal{D}$.  The specific construction of each observation (i.e., whether the observation contains a single explanation vs. an aggregation of explanations) depends on whether the use case is a global problem (using explanations to answer a question about the entire model) or a local problem (using an explanation to answer a question about the model's behavior at a particular point). For global problems, the use case label is the \emph{same} for a given model as opposed to for local problems, the use case label varies depending on the data-point of interest. A longer discussion can be found in the Appendix~\ref{appdx:expencoding}.

\textit{Multiple datasets $\mathcal{D}_i$ and models $f_i$.}  Instead of building each observation using a single base dataset and base model, for some use cases we sample each observation $i$ from a \emph{different} base dataset $\mathcal{D}_i$ (and use it to train a new base model $f_i$).  We do so because (a) global problems require multiple models to have a diverse set of use case labels
and (b) it may be undesirable for the agent to overfit to a single base dataset. 

\textit{Human Factors.}  When selecting explanation methods for evaluation, it is important to consider factors that influence the way humans reason about explanations. We advocate for researchers to select explanations that also satisfy desirable proxy metrics such as faithfulness and stability. Additionally, it is important to select explanations that are suitable for human use. 
We discuss these considerations in the context of our experiments and user study in more detail in Appendix~\ref{appdx:humanfactors}.

\textbf{Step 2: Agent training. } 
After generating a dataset, the researcher trains an algorithmic agent (e.g., a machine learning model) to predict the use case label for a given observation. While this step can be viewed as equivalent to training an ML model (i.e., requiring standard hyperparameter search and tuning), this step is also analogous to ``training'' phases in human subject studies in which humans are provided with labeled examples to convey relevant knowledge for the use case (e.g., presented example explanations and expected conclusions). 
The training process allows the algorithmic agent to implicitly learn relevant domain knowledge about the use case from labeled examples, even when the researcher does not have any knowledge a priori of how to predict the use case labels.

In terms of model architecture selection, the researcher should select an agent model that is able to encode and learn meaningful representations of the observations that are input to the agent (which include the data-point and explanations). For example, the researcher needs to define the encoding for the choice of explanation. There are some standard ways to encode certain types of explanations (e.g., a feature attribution explanation for tabular data can be encoded as a vector of length $d$ if $\vx \in \mathbb{R}^d$ and concatenated to the data-point to create an observation vector of length $2d$). However, in general, the researcher can leverage their domain knowledge in selecting the model: for example, a CNN architecture is appropriate for a base dataset containing images. We discuss in Appendix~\ref{appdx:expencoding} how we encode the explanations evaluated in our experiments.


\textbf{Step 3: Agent evaluation. } 
After training the agent, the researcher evaluates the agent's ability to make predictions on unseen test examples. This phase is analogous to the evaluation portion of typical user studies, where humans are tested on new examples. The agent's accuracy on these examples is a measure of the predictiveness of the information content for the use case: a relative higher accuracy is evidence that a particular type of information content is more useful. We note that the agent's performance (i.e. test set accuracy) is \emph{not} intended to be interpreted as the exact, anticipated human subject's performance when given the same set of information content as the algorithmic agent.  We discuss several factors beyond the predictive information in the explanation that may affect human performance in Appendix~\ref{appdx:humanfactors}. 
As such, researchers should focus on \emph{statistically significant differences} in agent performance when interpreting their SimEvals, as smaller differences between information content may not generalize to human performance.


\vspace{-0.25cm}
\section{Instantiating \simevals}\label{localexps}
\vspace{-0.25cm}

We instantiate our general framework to study the utility of popular post-hoc and by-design explanations (e.g., LIME, SHAP, Anchors, GAM) for three different use cases, spanning a diverse set of problems (forward simulation~\citep{hase_evaluating_2020}, counterfactual reasoning~\citep{kumar2020problems}, and data bugs~\citep{interpretinginterp}). The first two use cases are local problems that focus on base model behavior at a specific point, while the third is a global problem that asks questions about the base model behavior more broadly. The human subject evaluations that studied these use cases primarily focused on studying \emph{which explanation method} best helps users answer the use case questions.  


Next, we focus on a key portion of the \simeval{} framework that must be carefully defined, discussing the data generation process in Step (1) for each use case. Further details on data generation algorithms and agent architectures used for each use case can be found in Appendices~\ref{appdx:algos} and~\ref{appdx:expencoding}.

\textbf{Forward simulation}  measures whether a user correctly simulates (i.e., predicts) the ML model's prediction on a given data-point~\citep{poursabzi2018manipulating, chandrasekaran2018do, hase_evaluating_2020}. Our particular set-up is motivated by Hase and Bansal~\citep{hase_evaluating_2020}, where the authors test whether providing users with explanations helps users better predict a base model's behavior when compared to a control setting in which users are not given explanations. Here, each observation given as input to the \simeval{} agent consists of a single data-point and explanation $(\vx_i, E(\vx_i, f))$, where $\vx_i$ is sampled from the base dataset $\mathcal{D}$. Following the same procedure as Hase and Bansal, we split $\mathcal{D}$ into train, test, and validation sets.  The agent is trained on a dataset of observations from the training set and is then evaluated on observations 
from the unseen validation set. The use case label $u_i = f(\vx_i)$ is the model $f$'s output. 



\textbf{Counterfactual reasoning} is a concept from philosophy used to describe ``what if'' conditionals that are contrary to fact \citep{watcher2017counterfactual}.  Counterfactual reasoning can enable individuals to reason about how to achieve a desired outcome from a prediction model.  We study a specific instantiation of counterfactual reasoning from Kumar et al.~\citep{kumar2020problems}.  Our use case examines whether a model's output $f(\vx_i)$ increases for some input $\vx_i$ in a counterfactual setting where a given feature of $\vx_i$ is increased. Like forward simulation, each observation used to train and evaluate the agent consists of a single data-point but differs in that each observation also includes the base model prediction on that data-point: $(\vx_i, f(\vx_i), E(\vx_i, f))$. For a given input $\vx_i$, the use case label $u_i$ answers the question: ``Does $f(\vx_i)$ increase if a particular feature of $\vx_i$ is increased?''. We use the same agent model architecture as in the forward simulation use case.

One important difference for counterfactual reasoning is that each base dataset $\mathcal{D}_i$ and its corresponding base model $f_i$ can differ across observations.  We construct the agent's observations in this way to prevent the agent from simply memorizing where an individual base model is increasing versus decreasing with respect to a particular feature, instead encouraging the agent to learn a heuristic for this use case that generalizes across different base models (further discussion provided in Appendix~\ref{appdx:cfoverview}). 
\textbf{Data bugs} is a use case where the user is asked to predict whether the base model was trained on a dataset containing a systematic error. For this use case, the researcher additionally needs to provide a ``bug definition" $B$, which corrupts the given base dataset. We consider bug definitions studied in Kaur et al.~\citep{interpretinginterp} and summarize their motivation for identifying each of these bugs: The \textit{missing value} bug occurs when missing values for a feature are imputed using the mean feature value. This can bias the prediction model towards the mean value. The \textit{redundant feature} bug occurs when the dataset contains features that are redundant, or contain the same predictive information. This can distribute the true importance of the measure between the redundant features, failing to convey the measure's actual importance.  We note that while explanations may not be the only approach to identify these data bugs, the authors demonstrate that model explanations \emph{can} help users with this use case.



A distinguishing feature of the data bugs use case is that the study participants in Kaur et al.~\citep{interpretinginterp} were given a \emph{set} of data-points and explanations per observation because it might be easier to detect a model bug if given more than one data-point (an illustration provided in the Appendix~\ref{missingvalues}). 
As such, each observation for the agent is also a set of data-point, model prediction, and explanation tuples: $\{(\vx_i^{(j)}, f(\vx_i^{(j)}), E(\vx_i^{(j)}, f_i))\}_{j = 1}^S$, where $\vx_i^{(j)}$ denotes the $j$th element in the set which is part of the $i$th observation. 
In our experiments, we vary the proportion of datasets $\mathcal{D}_i$ that have bug $B$. For the datasets that have the bug, we set $u_i = 1$ and $u_i = 0$ otherwise.

For data bugs, each base dataset $\mathcal{D}_i$ and its corresponding base model $f_i$ differ across observations because data bugs is a \emph{global problem}, i.e. the use case label $u_i$ is a property of base model $f_i$. Specifically, we derive many $\mathcal{D}_i$ by subsampling one dataset $\mathcal{D}$. This allows us to create a variety of training (and evaluation) examples for the agent with varying use case labels, encouraging the agent to learn a heuristic that generalizes better across different sets of points. 
\section{Comparing Simulation and Humans}\label{experiments}

To demonstrate that \simevals{} are able to identify promising explanations, we run \simevals{} on the three use cases discussed in Section~\ref{localexps} and compare the algorithmic agent results to user study results, which come from either prior user studies or our own Mechanical Turk studies. 
Specifically, we validate that the algorithmic \simeval{} agent performs relatively well when given an explanation that a human can perform well with, and performs relatively worse when given an explanation that is not useful to a human. 
We include training details for our \simevals{} in Appendix~\ref{appdx:expdetails},  extended results (with error bars) and comparisons with human results in Appendix~\ref{appdx:full_results}. 

\subsection{User study details}

In our experiments, we compare the algorithmic \simeval{} agent's performance to the average human subject performance at predicting the use case label in a user study.  For the \textbf{Forward Simulation} use case, we compare the algorithmic agent's performance to human performance as reported in a previously conducted independent user study by Hase and Bansal~\citep{hase_evaluating_2020}.  We conduct new user studies to measure human performance on two of the use cases.  For \textbf{Counterfactual Reasoning}, there is no existing user study from the motivating work from Kumar et al.~\citep{kumar2020problems}, necessitating that we run our own.  For \textbf{Data Bugs}, we conducted our own study that, unlike the original one by Kaur et al~\citep{interpretinginterp}, individually measures the usefulness of candidate explanations.
We expand on the differences between the two studies in Section~\ref{sec:databugsexp}.


Our user studies were conducted on Amazon Mechanical Turk, where we presented Turkers with the observations sampled from the dataset that \simevals{} were trained and evaluated on. We choose a between-subjects study design, where each participant is randomly assigned to one explanation settings from the \simevals. In non-baseline settings, participants are also provided with explanations for each $\vx_i$, which we described to the Turkers as importance scores (Figure \ref{fig:turk-instructions}).
The study is split into two parts. First, each participant completes a ``Train'' survey (analogous to agent training) where they are asked to predict the use case label for 15 observations and provided feedback on whether their responses were correct after every 5 observations. Then participants complete a ``Test'' survey (analogous to agent evaluation) where they predict the use case label for another 15 observations without feedback. 
Each question corresponds directly to 1 randomly sampled observation used to train or test the algorithmic agent.  For each use case, we recruit 80 total participants: 20 participants for each explanation setting. Other details, including the study interface and instructions for each use case, participant recruitment process, details of initial pilot studies are provided in Appendix~\ref{appdx:mturk}.

\subsection{Forward Simulation}




Following Hase and Bansal~\citep{hase_evaluating_2020}, we compare the prediction accuracy of our agent in the `No explanation' baseline setting (where the agent/human receives only the data-point without explanation) 
with its prediction accuracy when given both the data-point and its corresponding explanation. 
The authors evaluate 5 explanations in their user study: 2 are well-known explanations (LIME~\citep{ribeiro2016should}, Anchors~\citep{ribeiro2018anchors}) and the rest are bespoke. In this work, we do not evaluate the authors' bespoke methods since none outperformed the standard methods as described in Appendix~\ref{sec:summary}. 

\begin{table}[h!]
\centering
\caption{(Left) Test accuracy of the agent model (averaged across $3$ random seeds) when varying the number of training observations the agent receives to perform forward simulation, recreating trends in average human subject test accuracy (Right) from Hase and Bansal~\citep{hase_evaluating_2020}.  
}
\begin{tabular}{lllll|c}
\hline
                       & \multicolumn{4}{c|}{\begin{tabular}[c]{@{}c@{}}Agent Test Accuracy\\ (Varying Train Set Size)\end{tabular}} & \begin{tabular}[c]{@{}c@{}}Human Test Accuracy\\  from~\citep{hase_evaluating_2020}\end{tabular} \\
Explanation            & 16                        & 32                        & 100                      & 1000                     & \multicolumn{1}{l}{}                                           \\ \hline
LIME            & 94.2\%                    & 99.8\%                    & 100\%                    & 100\%                    & 81.9\% $\pm$ 8.8\%                                                      \\
Anchors       & 89.2\%                    & 93.5\%                    & 94.7\%                   & 93.7\%                   & 75.8\% $\pm$ 8.5\%                                                        \\
No explanation & 82.3\%                    & 83.7\%                    & 85.7\%                   & 88.7\%                   & 70.7\% $\pm$ 6.9\%                                                        \\
\hline
\end{tabular}
\label{table:forwardsim}
\end{table}


\textit{Findings:} Hase and Bansal~\citep{hase_evaluating_2020} found that participants who were given the LIME explanation for forward simulation achieved an accuracy increase that is more than twice that achieved by other methods. As shown in Table~\ref{table:forwardsim}, we find that \simevals{} corroborates this result from the user study even when varying the agent's train set size. In Appendix~\ref{sec:fwd_sim_results}, we provide additional \simevals{} which hypothesize that SHAP and GAM are also promising choices for this use case.


\subsection{Counterfactual reasoning} \label{sec:userstudy}
The counterfactual reasoning use case is introduced by Kumar et al.~\citep{kumar2020problems}, where the authors discuss why SHAP feature importance scores ~\citep{lundberg2017unified} do not explicitly attempt to provide guidance for the use case. Specifically, a positive SHAP feature importance score does not necessarily imply that increasing that feature will increase the model's prediction.  We additionally hypothesize that an approximation-based explanation like LIME might be more useful for this use case than SHAP. We conduct a user study for this use case since Kumar et al. did not conduct a user study. 

\begin{table}[h!]
\centering
\caption{Average agent test set accuracy (Left) aligns with average Turker test survey accuracy (Right) for the counterfactual reasoning use case. (Left) Agent test set accuracies as we vary the number of training observations the agent receives to perform counterfactual reasoning. (Right) The 95\% confidence interval of average user accuracy on 15 test observations where for each explanation setting we recruited $N=20$ Turkers.}
\begin{tabular}{lcccc|c}
\hline
               & \multicolumn{4}{c|}{\begin{tabular}[c]{@{}c@{}}Agent Test Accuracy \\ (Varying Train Set Size)\end{tabular}} & \begin{tabular}[c]{@{}c@{}}Human Test \\ Accuracy\end{tabular} \\
Explanation    & 4                         & 16                        & 100                       & 1000                     &                                                                \\ \hline
LIME           & 92.9\%                    & 94.5\%                    & 99.2\%                    & 99.7\%                   & 69.4\% $\pm$ 12.4\%                                            \\
SHAP           & 56.3\%                    & 52.5\%                    & 57.3\%                    & 64.9\%                   & 41.4\% $\pm$ 6.14\%                                            \\
GAM            & 56.0\%                    & 53.5\%                    & 57.3\%                    & 63.5\%                   & 45.7\% $\pm$ 7.19\%                                            \\
No explanation & 52.1\%                    & 55.6\%                    & 57.9\%                    & 60.3\%                   & 48.6\% $\pm$ 5.61\%                                            \\ \hline
\end{tabular}
\label{table:cf-results}
\end{table}


\textit{Findings:} As shown in Table \ref{table:cf-results} (Left), we find that the simulated agent had significantly higher accuracy when given LIME explanations. Additionally, we find that this trend is consistent across all training set sizes: the agent achieves a 37\% increase in validation accuracy when given LIME instead of any other explanation after observing only 4 train observations. These results suggest that, of the set of explanations considered, LIME is the most promising explanation for the counterfactual reasoning use case. In Appendix~\ref{appdx:ablation} we demonstrate that LIME outperforms all other explanation methods consistently across different types of model architectures. 

As shown in Table~\ref{table:cf-results} (Right), the gap in agent performance between LIME and the other explanations is reflected in the human test accuracy in our Turk study. In particular, a number of Turkers are able to use the predictiveness of LIME to complete the use case with high accuracy; whereas, in general, Turkers using SHAP, GAM, and No Explanation perform no better than random guessing. We ran an ANOVA test and found a statistically significant difference between the Turkers' accuracy using LIME vs. all other explanation settings and no statistically significant difference between the other settings (Appendix~\ref{appdx:cfresults}). Note that there is also significant overlap in the error bars of agent performance on SHAP, GAM, No Explanation (Table~\ref{table:cfreasoning}). We observe greater variance in participant accuracy between participants for LIME compared to the other explanations. We believe that this is because many, but not all, Turkers were able to learn how to use the LIME explanation successfully. 


\subsection{Data bugs}\label{sec:databugsexp}
Kaur et al.~\citep{interpretinginterp} study two explanations (SHAP \citep{lundberg2017unified} and GAMs \citep{hastie1987generalized}) on two types of data bugs (missing values and redundant features). In their study, users were prompted to suggest bugs that they believed may be present via semi-structured explorations using these explanation tools. They found that 4 out of 11 users mentioned a missing values bug and 3 users mentioned redundant features. Their results suggest that both SHAP and GAM may be useful for finding these types of data bugs, though they do not individually test each explanation's usefulness and they also did not consider providing users with baseline information content (without model explanations) to compare with settings where users were presented with explanations. Thus, for this use case, we conducted our own Turk user study on the missing values bug, evaluating both the explanations considered in the original study and including additional explanation/baseline comparisons.
We present \simeval{} results for the missing values bug below (redundant features can be found in Appendix~\ref{appdx:redfeat}).






\textbf{Missing Values:} Kaur et al.~\citep{interpretinginterp} implement the missing values bug by replacing the `Age' value with the mean value of 38 for 10\% of adults with $>50$k income. We detail how we recreate this bug in Appendix~\ref{missingvalues}. In addition to comparing various explanation methods, we also use \simevals{} to evaluate the effect of varying the number of data-points per train set observation (e.g. the size $S$ of each set). We note is another type of information content that the researcher must select when running a user study.

\begin{table}[h!]
\centering
\caption{Average agent test set accuracy (Left) aligns with average Turker test survey accuracy (Right) for the data bugs (missing values) use case. (Left) Agent test set accuracies as we vary the number of training observations the agent receives to perform counterfactual reasoning. (Right) The 95\% confidence interval of average user accuracy on 15 test observations where for each explanation setting we recruited $N=20$ Turkers. 
We denote explanation settings that were considered in the original user study with $\star$.\\}
\begin{tabular}{lllll|l}
\hline
                 & \multicolumn{4}{c|}{\begin{tabular}[c]{@{}c@{}}Agent Test Accuracy\\ (Number of Data-points Per Observation)\end{tabular}} & \multicolumn{1}{c}{\begin{tabular}[c]{@{}c@{}}Human Test\\ Accuracy\end{tabular}} \\
Explanation      & 1                             & 10                           & 100                          & 1000                         & \multicolumn{1}{c}{}                                                          \\ \hline
SHAP $\star$     & 63.2\%                        & 84\%                         & 99.8\%                       & 100\%                        & 67.4\% $\pm$ 11.5\%                                                                             \\
GAM  $\star$     & 64.8\%                        & 87.7\%                       & 100\%                        & 100\%                        & 64.4\% $\pm$ 7.2\%                                                                               \\
LIME             & 55.2\%                        & 56.9\%                       & 64.5\%                       & 75.9\%                       & 48.0\% $\pm$ 5.4\%                                                                              \\
Model Prediction & 57.5\%                        & 57.4\%                       & 58.2\%                       & 67.3\%                       & 40.7\% $\pm$ 5.2\%                                                                             \\ \hline
\end{tabular}
\label{missingvaluesvary}
\end{table}

\textit{Findings:} As shown in Table~\ref{missingvaluesvary} (Left), we verify using \simevals{} that SHAP and GAM explanations contain predictive information to identify the missing values bug. 
We also find that the accuracy of the algorithmic agent drops when given too few data-points (i.e., for observation sets of size $\leq 10$).  This means that a small number of data-points do not contain sufficient signal to detect the bug. Since the researcher would need to present a large number of data-points and explanations to a human user in a human-interpretable manner, a user interface is needed where the explanations are presented in an aggregated fashion, which is indeed the approach taken by both Kaur et al.~\citep{interpretinginterp} and our study.
In Appendix~\ref{appdx:full_results}, we observe similar trends in the relative performance of explanation methods even when varying the bug strength and the parameterization of the agent model.

In our Turk study, we utilized the findings from our \simevals{} experiments and presented the explanations for $S = 1000$ data-points per observation in an aggregated fashion (via scatter plots as shown in Figure~\ref{fig:turk-q-databugs}). As shown in Table~\ref{missingvaluesvary} (Right), participants assigned to SHAP or GAM consistently outperformed the participants assigned to LIME or the Model Prediction baseline, even though our participant population for the user study were dissimilar from the participant population (data scientists) of the original user study. We conducted pairwise statistical analyses and found statistical significance between SHAP and LIME/Model Prediction as well as GAM and LIME/Model Prediction and no statistical significance between other pairs of experimental conditions. 

\vspace{-0.25cm}
\section{Discussion} \label{sec:discussion}
\vspace{-0.35cm}

Our experiments and analyses in Section~\ref{experiments} demonstrate that \simevals{} are able to provide an informative measure of the predictive information contained in a design set-up, and as such can identify promising explanations from a set of candidate explanations.  
However, we emphasize that because \simevals{} only measure predictive information, the algorithmic agent's accuracy \emph{cannot} be directly interpreted as the explanation's utility to a human, as evidenced by the differences between raw agent and human test accuracies in Tables~\ref{table:forwardsim}, \ref{table:cf-results}, and~\ref{missingvaluesvary}.
Furthermore, we note that small differences between \simeval{} agent's relative performance on explanations (e.g., in Table~\ref{table:cf-results}) may not generalize to humans.
We discuss potential human factors that may contribute to these differences between \simeval{} and human performance in Appendix~\ref{appdx:humanfactors}. These findings suggest an important direction for future research to better understand these differences between agent and human, i.e.,  how human factors beyond predictive information affect humans' ability to reason about model explanations.

\vspace{-0.25cm}
\section{Conclusion}
\vspace{-0.35cm}
We proposed a use-case-grounded algorithmic evaluation called \simevals{} to efficiently screen choices of information content and identify good candidates for a user study. Each \simeval{} involves training an agent to predict the use case label given the information content that would be presented to a user. 
We instantiated \simevals{} on three use cases motivated by prior work, and demonstrated that the agent's test set accuracy can be used as a measure of the predictiveness of information content (more specifically, of an explanation method) for each use case. 
We found that humans perform significantly better on explanations that \simevals{} selects as promising compared to other explanations.
While our experiments focus primarily on the design choice of selecting explanation methods, we note that researchers can also use \simevals{} to screen the predictiveness of other types of information content, such as hyperparameter selection for explanation methods or the baseline information given with each observation. We hope that this work can be incorporated into future explanation evaluation workflows to enhance the efficiency and effectiveness of user study design.

\section{Acknowledgements}

We are grateful for helpful feedback from Chirag Agarwal, Zana Bucinca, Alex Cabrera, Lucio Dery, Elena Glassman, Joon Sik Kim, Satyapriya Krishna, Isaac Lage, Martin Pawelcyzk, Danish Pruthi, Rattana Pukdee, Isha Puri, Junhong Shen, Manuela Veloso.

This work was supported in part by the National Science Foundation grants IIS1705121, IIS1838017, IIS2046613, IIS2112471, an Amazon Web Services Award, a Facebook Faculty Research Award, funding from Booz Allen Hamilton Inc., and a Block Center Grant. Any opinions, findings and conclusions or recommendations expressed in this material are those of the author(s) and do not necessarily reflect the views of any of these funding agencies.

\bibliographystyle{plain}
\bibliography{ref}

\newpage
\appendix

\section{Examples of algorithmic agents in prior work} \label{appdx:existingwork}

We can unify existing ad-hoc examples that use algorithmic agents to perform use case grounded evaluations under our framework.  We group existing algorithmic evaluations into two categories: Heuristic-Based and Learning-Based.

\subsection{Heuristic-Based Evaluations.}

These algorithmic agents were not trained, and instead were defined using hand-designed heuristics. In contrast, our agent is trained on observations and use case labels. 

\textbf{Expo}~\citep{expo}: The use case studied is whether an explanation can be used to determine how to modify features of an instance to achieve a target prediction. There is no agent training process: the agent uses a randomized greedy heuristic (which involves picking the largest feature). The agent is evaluated by the number of modifications that it makes to reach the target prediction. 

\textbf{Influence functions}~\citep{koh2017understanding} \textbf{/ RPS}~\citep{rps}: The use case studied is whether explanations can be used to identify mislabeled examples in a dataset. The agent is also heuristic-based, flagging points with the largest influences to be inspected for potentially being mislabeled. The agent is evaluated by the number of correct points it selects for inspection.

\textbf{Anchors}~\citep{ribeiro2018anchors}: The use case studied is whether explanations can help a user make accurate predictions on unseen instances. The authors specify a heuristic that they use to compare Anchors with LIME. The heuristic for Anchors is simply to check whether the conditions are satisfied. The heuristic for LIME is to use the linear approximation directly (while additionally checking whether the point of prediction is near the approximation point). The agent is evaluated using its precision when used to predict labels on unseen instances.

\subsection{Learning-Based Evaluations.}

\textbf{Student-teacher Models} \cite{pruthi2020evaluating}:  The use case studied is whether explanations can help a student model learn to simulate a teacher model for a sentiment analysis task and question answering task.  The authors quantify the benefit of an explanation as the improvement in simulation accuracy when the agent has access to explanations during the training process versus when the agent does not. One important distinction from our approach is that it is noted in \citep{pruthi2020evaluating} that their algorithmic evaluation does \emph{not} attempt to replicate the evaluation protocols used in a human subject study. This point is affirmed by the significant difference between the results of the authors' human subject study vs. their algorithmic evaluation. In contrast, our approach is designed with a human subject study in mind: we explicitly do attempt to construct the agent's observations in a way that reflects the information that would be presented to a human subject.  

Other work \citep{ebrahimi2021remembering, viviano2021saliency} similar to \citep{pruthi2020evaluating} that evaluates explanations without a human in the loop typically examines how explanations can improve the model training process. In contrast, our work focuses on use cases where explanations are intended to be shown to and used by humans. 

\subsection{Distinction from prior work}

In Table~\ref{tab:priorwork}, we summarize prior work along two main axes: their algorithmic evaluation of explanations (if they conducted them) and their human evaluation of explanations (if they conducted them). On the algorithmic evaluation front, no prior work allows their agent to learn how to use explanations (i.e., a researcher does not need to specify a heuristic for how the explanation will be used) \emph{and} provides a general framework that encompasses many use cases. Furthermore, no prior work extensively verifies their evaluation framework with human evaluation to show the comparability between algorithmic and human performance, whereas our work evaluates as many explanations as many user studies. 

\begin{table}[h!]
\begin{tabular}{|l|ll|lll|}
\hline
                                                                          & \multicolumn{2}{c|}{Algorithmic Evaluation}                                                          & \multicolumn{3}{c|}{Human Evaluation}                                                                                                                                            \\ \hline
                                                                          & \multicolumn{1}{c|}{\begin{tabular}[c]{@{}c@{}}Learning-based? \\ (i.e.,no heuristic\\ required).\end{tabular}} & \begin{tabular}[c]{@{}l@{}}Does framework \\ generalize?\end{tabular} & \multicolumn{1}{c|}{\begin{tabular}[c]{@{}c@{}}Conducts a  \\ user study?\end{tabular}} & \multicolumn{1}{c|}{\begin{tabular}[c]{@{}c@{}}Agent matches \\ human?\end{tabular}} & \multicolumn{1}{c|}{\begin{tabular}[c]{@{}c@{}}Num. explanations \\ (baselines) evaluated\end{tabular}} \\ \hline
Expo                                                                      & \multicolumn{1}{l|}{No}              & No                                                            & \multicolumn{1}{l|}{Yes}         & \multicolumn{1}{l|}{No}             & 1 (1)                                                                                                   \\ \hline
IF/RPS                                                            & \multicolumn{1}{l|}{No}              & No                                                            & \multicolumn{1}{l|}{No}          & \multicolumn{1}{l|}{–}              & –                                                                                                       \\ \hline
S-T Models                                                             & \multicolumn{1}{l|}{Yes}             & No                                                            & \multicolumn{1}{l|}{Yes}         & \multicolumn{1}{l|}{No}             & 2 (1)                                                                                                   \\ \hline
Anchors                                                                   & \multicolumn{1}{l|}{No}              & No                                                            & \multicolumn{1}{l|}{Yes}         & \multicolumn{1}{l|}{Yes}            & 2 (0)                                                                                                   \\ \hline
\begin{tabular}[c]{@{}l@{}}User studies\end{tabular} & \multicolumn{1}{l|}{–}               & –                                                             & \multicolumn{1}{l|}{Yes}         & \multicolumn{1}{l|}{–}              & $\sim$2-4 (1-2)                                                                                          \\ \hline
\textbf{\begin{tabular}[c]{@{}l@{}}Ours \\ (SimEvals)\end{tabular}}       & \multicolumn{1}{l|}{Yes}             & Yes                                                           & \multicolumn{1}{l|}{Yes}         & \multicolumn{1}{l|}{Yes}            & 4 (2)                                                                                                   \\ \hline
\end{tabular}
\label{tab:priorwork}
\end{table}

\section{Use Case Algorithms}\label{appdx:algos}

We provide more detailed algorithms to describe how the dataset is generated for the forward simulation and counterfactual reasoning use cases. 

\subsection{Forward simulation}

\begin{itemize}
    \item Given base dataset $\mathcal{D}$, split into train/test/validation
    \item Train model $f$ on the train set
    \item Let $N_T$ be size of desired training dataset and $N_V$ be size of the desired validation dataset
    \item Sample $N_T$ points from the train set and for each point $\vx_i \in \mathcal{D}_{train}$ generate an explanation of $f$ at that point, add the tuple $(\vx_i, E(\vx_i,f))$ to the training dataset.
    \item Repeat process for $N_V$ points but from the validation set
\end{itemize}

Note the baseline condition would be generated in the same way, but excluding the explanation.

\subsection{Counterfactual Reasoning}

\begin{itemize}
    \item Given base model family $\mathcal{F}$, desired dataset size $N$
    \item For $i = 1, ..., N$:
    \begin{itemize}
        \item Sample saddle-point function $g_i$ as detailed in Appendix \ref{apdx:cf-synthetic-data}
        \item Train $f_i$ on $g_i$ as detailed in Appendix \ref{apdx:cf-synthetic-data}
        \item Sample new point $\vx_i$ from $g_i$
        \item Get prediction $f_i(\vx_i)$
        \item Generate $E(\vx_i, f_i)$ of $f_i$ at $\vx_i$
    \end{itemize}
    \item Set the $i$th observation to be: $ (\vx_i, f_i(\vx_i), E(\vx_i, f_i))$
\end{itemize}

\subsection{Data Bugs}

\begin{algorithm}
  \caption{generates a dataset of $N$ observations for the data bug use case.}
  \begin{algorithmic}[1]
    \STATE Given base dataset $\mathcal{D}$, base model family $\mathcal{F}$ and bug specification $B$
    \STATE Let $N$ be the size of desired dataset, \\
    $S$ be the observation set size
    \FOR{$i$ in $1...N$} 
        \STATE {Subsample $\mathcal{D}_i$ from $\mathcal{D}$}
        \STATE {With probability 0.5, apply $B$ to $\mathcal{D}_i$}
        \STATE {Train $f_i$ on $\mathcal{D}_i$}
        \FOR{$j$ in $1...S$}
            \STATE Sample $\vx_i^{(j)}$ from $\mathcal{D}_i$
            \STATE Get prediction $f_i(\vx_i^{(j)})$ 
            \STATE Generate $E(\vx_i^{(j)},f_i)$ of $f_i$ at $\vx_i^{(j)}$ 
        \ENDFOR
        \STATE {Set the $i$th observation to be: \\$\{(\vx_i^{(j)}, f_i(\vx_i^{(j)}), E(\vx_i^{(j)}, f_i))\}_{j = 1}^S$}
    \ENDFOR
  \end{algorithmic}
 \label{algo:databugs}
\end{algorithm}

\section{Summary of prior user studies}\label{sec:summary}

We discuss the findings from the prior user studies that we recreate results from.

\textbf{Forward Simulation} from~\citep{hase_evaluating_2020}. The authors evaluate 5 explanations in this paper: 2 (LIME, Anchors) are well-known, open-source explanations and 3 (Prototype, DB, and Composite) are bespoke explanations. In the tabular setting, the authors find that LIME outperforms all other explanations (change of 11.25\% in user accuracy compared to Anchors 5.01\%, Prototype 1.68\%, 5.27\%, 0.33\%). The base accuracy before explanations was 70.74\%. Since none of the bespoke explanations outperformed the open-source explanations, we chose not to re-implement them in our study.

\textbf{Data Bugs} from~\citep{interpretinginterp}. The authors evaluate 2 explanations: SHAP and GAM. The set-up of this study did not explicitly ask the users to answer whether the bug was present, but rather asked the user to identify the bugs themselves (i.e., if a user could identify the existence of the bug, then the explanation was useful). The authors find that 4 out of 11 data scientists were able to identify the missing values bug and 3 out of 11 the redundant features bug. While the authors do not distinguish when SHAP is helpful versus GAM. From Figure 1 of their paper, SHAP and GAM seem to have similar visualizations and thus might equally help a user. We find similar results with our simulated agent.




\section{Counterfactual Reasoning Use Case Details} \label{appdx:cfoverview}

\subsection{Distinction from Counterfactual Explanations}\label{apdx:cf-distinction}

There are significant differences between our use case and the desiderata for which counterfactual explanation methods for ML models are developed. 


\paragraph{Why existing counterfactual explanation methods do not apply:}  A growing body of IML research has proposed new methods to provide counterfactual explanations for a black-box model~\citep{ustun2018actionable, mothilal2020explaining, karimi2021algorithmic}.  These methods provide explanations that fulfill the desiderata introduced in \citep{watcher2017counterfactual}, defined formally as:

\begin{definition}\label{defn:cf}
Given input $\vx$, model $f$, and desired outcome set $\mathcal{Y}$, counterfactual explanation $\vx'$ is the \emph{closest possible point} to $\vx$ such that the model's prediction on the counterfactual point is in the desired outcome set: $f(\vx') \in \mathcal{Y}$.  
\end{definition}

The real-world utility of counterfactual explanations relies on a large number of assumptions \citep{barocas2020hidden}.
Many variants of the counterfactual problem have been proposed to account for a wide range of real world objectives and constraints, such as the feasibility or difficulty of the proposed actions (recourse) \citep{ustun2018actionable}. Despite these varying goals, however, to our knowledge prior work on counterfactual explanations shares the general problem set-up in Definition \ref{defn:cf}.

While this problem set-up in Definition \ref{defn:cf} is \emph{related} to our use case, it is significantly \emph{different} for a number of reasons.  Primarily, while the counterfactual explanation $\vx'$ provided by these methods is encouraged to be ``close'' to the original input $\vx$, the counterfactual $\vx'$ can increase or decrease any subset of input features within the problem's constraints.  Thus, there may be a wide range of possible counterfactuals $\vx'$ of varying ``closeness'' to $\vx$ that result in the desired outcome $f(\vx')$.  In contrast, the use case asks a more specific question of what happens when a specific \emph{feature $i$} is \emph{increased} and all other features are held constant.  

In a scenario where it may be possible to achieve an increased model prediction $f(\vx')$ by changing other subsets of features excluding feature $i$, then the provided counterfactual $\vx'$ will give no information relevant to the use case.  Similarly, it may be possible that $f(\vx')$ may decrease when some feature $j \neq i$ is changed \emph{and} feature $i$ is increased, suggesting that $f$ decreases when feature $i$ is increased; but if all other features $j \neq i$ are held \emph{constant}, then $f$ \emph{increases} when feature $i$ is increased.  In these cases, the optimal counterfactual explanation $\vx'$ may give no information about how $f(\vx)$ varies with feature $i$. 

Due to this misalignment between the goals of existing counterfactual explanation methods and our particular use case, we exclude these counterfactual explanation methods from our evaluation.  We instead focus our evaluation on common feature attribution methods.  

\subsection{Synthetic Data}\label{apdx:cf-synthetic-data}

Inspired by Kumar et al.~\citep{kumar2020problems}, we construct a 2D toy base dataset comprising of points sampled from a saddle-point function.  We construct a low-dimensional dataset to reduce the complexity of the task given to MTurkers, as prior work~\citep{poursabzi2018manipulating} found that Turkers achieved higher accuracy when presented with data that had fewer features.  

Each data-point $(\vx_i: [x_{i, 1}, x_{i, 2}], y_i)$ used to train predictive model $f_i$ is sampled from its \emph{own} saddle-point function $g_i$ with parameters chosen stochastically (detailed further in Section \ref{sec:1-saddlepoint} below).  Because each observation is sampled from a different function, each observation also has its own predictive model $f_i$.  Importantly, we chose to sample points from a saddle-point function (with no noise) because saddle-point functions are either \emph{strictly concave} or \emph{strictly convex} with respect to each feature $x_{i, 1}, x_{i, 2}$.  Thus, if the predictive model $f_i$ correctly learns the saddle-point function $g_i$, then ``ground truth'' use case labels $u_i$ of whether or not the predictive model's output increases with feature $x_{i, 2}$ are available for all points.

\textbf{Why each point from a separate function?:} We chose to sample every observation given to the agent from its own separate saddle-point function to increase the difficulty of the agent prediction task. Consider for comparison an alternative naive dataset generation procedure where all observations $\{(\vx_i, f(\vx_i)), u_i\}_{i = 1}^N$ are sampled from the \emph{same} saddle-point function $g$ (and consequently share the same predictive model $f$).  Given enough observations $N$, because $f$ is smooth and either strictly concave or convex, an agent can memorize for which points  $(\vx_i, f(\vx_i))$ the function $f$ is increasing vs. decreasing with respect to feature $x_{i, 2}$.  Thus, the agent can effectively memorize the use case labels without needing to use explanations. This is undesirable as our intention in training the agent is to evaluate candidate explanation methods.

Therefore instead of using the same predictive model for all observations, we trained different predictive models on observations sampled from different saddle-point functions to (a) prevent the agent from learning a heuristic specific to any singular function and (b) to encourage the agent to learn a heuristic that will generalize well on new observations from different functions.

Below, we detail how data-points and use case labels are defined for each saddle-point function (Section \ref{sec:1-saddlepoint}) and how the agent's train and validation sets are constructed (Section \ref{sec:2-saddlepoint}).

\subsubsection{Saddle-Point Functions}\label{sec:1-saddlepoint}

For each observation $i$, we construct saddle-point function $g_i$ as follows:

First, critical points $x^*_1, x^*_2$ are chosen uniformly at random in range $[10, 15]$.  Second, an indicator variable $Z$ is sampled from a Bernoulli distribution with parameter $p = 0.5$. 

We define saddle-point function $g_i$ as:
\begin{equation}
    g_i(x_1, x_2) = (x_1 - x^*_1)^2 - (x_2 - x^*_2)^2
\end{equation}


To sample data-point pairs $(\vx_i: [x_{i, 1}, x_{i, 2}], y_i)$ and their corresponding ground truth use case labels $u_i$ from $g_i$:
\begin{itemize}
    \item $x_{i, 1}$ and $x_{i, 2}$ are sampled uniformly in the ranges $[x^*_1 \pm 5]$ and $[x^*_2 \pm 5]$ respectively.
    \item Generate outcomes $y_i$ as:
    \begin{equation}
    y_i =
    \begin{cases*}
      g_i(x_{i, 1}, x_{i ,2}) + 30, & if $Z = 1$  \\
      - g_i(x_{i, 1}, x_{i ,2}) + 30,       &  if $Z = 0$
    \end{cases*}
    \end{equation}
    
    \item Generate use case labels $u_i$ using indicator variables $\mathbbm{1}(\cdot)$ as:
    
    \begin{equation}
    u_i =
    \begin{cases*}
      \mathbbm{1}(x_{i, 2} \leq x_2^*), & if $Z = 1$  \\
      \mathbbm{1}(x_{i, 2} > x_2^*),      & if $Z = 0$
    \end{cases*}
    \end{equation}
    
\end{itemize} 

\begin{figure}[H]
\centering
  \includegraphics[scale=0.3]{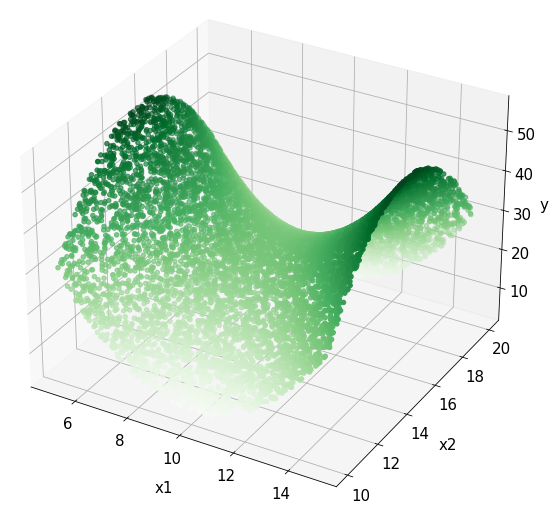}
  \caption{Scatterplot of $10,000$ points sampled from the saddle-plot function with $x_1^* = 10$, $x_2^* = 15$, and $Z = 1$. This function is \emph{strictly concave} with respect to feature $x_2$.}
  \label{fig:saddlepoint2d}
\end{figure}

Figure \ref{fig:saddlepoint2d} shows the 2D toy dataset generated using critical points $x_1^* = 10$, $x_2^* = 15$ and $Z = 1$.  Notice that because $y = g_i(x_1, x_2) + 30$ is strictly concave with respect to feature $x_2$, the function's output \emph{increases} as $x_2$ increases for all points where $x_2 \leq x_2^*$.  At the critical point of $x_2^*$,  the use case labels $u$ switch from $1$ to $0$:  for points where $x_2 > x_2^*$, then the function's output \emph{decreases} as $x_2$ increases. 

In contrast, when $Z = 0$, then data-points $y = - g_i(x_1, x_2) + 30$ are generated from a function that is the negation of Figure \ref{fig:saddlepoint2d} that is strictly convex with respect to feature $x_2$.  Thus the use case labels are the negation of the use case labels for $Z = 1$: they are now $1$ when $x_2 > x_2^*$.

\subsubsection{Running the Simulated Evaluation}\label{sec:2-saddlepoint}

Below we detail how predictive models $f_i$ and the agent's train and validation sets are constructed.

\textbf{Training predictive models $f_i$. }  For each observation $i$, we generate $N = 5000$ data-points in range $[x^*_1 \pm 5, x^*_2 \pm 5]$ following Section \ref{sec:1-saddlepoint} and train the Light-GBM Regressor $f_i$.  We validate by visual inspection that the learned models $f_i$ are ``smooth'', i.e. that the use case labels $u_i$ derived using the definition in Section \ref{sec:1-saddlepoint} are correct by visualizing the model's output $f_i(\vx_i)$ over the domain $\vx_i \in [x^*_1 \pm 5, x^*_2 \pm 5]$.

\textbf{Sampling observations $(\vx_i, f_i(\vx_i), u_i)$ to give to the agent.}  After a predictive model $f_i$ has been trained, we sample a single point $\vx_i$ from the fixed range $[11.25, 13.75]$ to construct observation $i$ for the agent. While this range is a \emph{subset} of the range in which the critical points $x_1^*, x_2^*$ and the training data points were sampled, we intentionally sample the input points $\vx_i$ in a more narrow range to increase the difficulty of the agent prediction task. Specifically, for the chosen range, the probability that the randomly selected critical point $x_2^*$ lies outside of this range is $50\%$, making it difficult for the agent to easily infer the use case label from the model prediction $f_i(\vx_i)$ and covariates $\vx_i$ alone.

Repeating this process several times, we construct a dataset of $10,000$ different observations (e.g. we train $10,000$ different models $f_i$ corresponding to $10,000$ different saddle-point functions $g_i$).  For the 4 different explanation settings, we use each observation's respective trained predictive model $f_i$ to generate explanations. To train the agent, we split these $10,000$ observations into a train, test, and validation set.

\subsection{Limitations \& Future Work}

We note that there are a few limitations to our use case definition and evaluation.  First, our use case definition does not account for the feasibility or cost of increasing the specific feature under study, which may be of practical importance in a real-world application.  Second, we note that it may be difficult or subjective to derive ground-truth use case labels of whether or not a predictive model's outcome will increase locally if a feature is increased for a non-convex prediction model $f$. The naive approach of taking the gradient of $f$ at point $\vx$ may not be representative of the behavior of function $f$ in a region around $\vx$ \citep{smilkov2017smooth}.  In practice, we recommend that ground-truth use case labels are defined empirically by averaging the model's predictions in a local region around input $\vx$ using expert knowledge to define a meaningful and representative local region.  Third, we note that to our knowledge, there is no explanation method that is explicitly developed to address this use case, which is an open direction for future work.

\section{Explanation Encoding} \label{appdx:expencoding}

\subsection{Global vs. Local}

The number of data-points included in each observation can vary depending on the \emph{explanation type} (global vs. local explanations) and \emph{problem type} (global vs. local problems).  Table \ref{table:observation_problem} summarizes how we recommend constructing each observation for each combination of problem and explanation type.  We elaborate on several of our choices below.


\begin{table}[H]
\caption{Describes how to construct each observation $i$ using global and local explanations for both global and local problems. Notation: $S$ is the size of the observation set; $\mathcal{D}$ denotes the base dataset; $f$ denotes the base model; $\vx_i$ denotes a point sampled from the base dataset.  Note that each observation $i$ for global problems by definition must use a \emph{different} dataset $\mathcal{D}_i$ and model $f_i$; vs. local explanations may use the same model $f$ and dataset $\mathcal{D}$ (elaborated on below).\\}
\centering
\begin{tabular}{l|c|c|}
\cline{2-3}
                                         & Global Problem        & Local Problem \\ \hline
\multicolumn{1}{|l|}{Global exp} & $E(f_i,\mathcal{D}_i)$                & $E(f,\mathcal{D})$            \\ \hline
\multicolumn{1}{|l|}{Local exp}  & $\{E(\vx_i^{(j)}, f_i)\}_{j = 1}^S$ for $\vx_i^{(j)}$ in $\mathcal{D}_i$ & $E(\vx_i,f)$        \\ \hline
\end{tabular}
\label{table:observation_problem}
\end{table}

\begin{itemize}
    \item \textbf{Global Problem + Global Explanation: }  As discussed in Section \ref{generalframework}, global problems require that each observation uses a different base predictive model $f_i$ because the use case $u_i$ labels are a property of the model (and thus would be the same for all observations if they all shared the same model).  Therefore each observation $i$ contains the global explanation for that model $f_i$.
    \item \textbf{Global Problem + Local Explanation: } Global problems require that each observation uses a different predictive model, but local explanations only explain the model's prediction on a single data-point.  A single data-point may not contain enough information to infer a global property of predictive model $f_i$.  To address this problem, the agent can take \emph{multiple} (e.g. a set of $S$) data-points and explanations per observation.
    \item \textbf{Local Problem + Global Explanation: }  A local problem defines a different use-case label for each individual data-point $\vx_i$.  However, each base model $f$ and base dataset $\mathcal{D}$ only has 1 global explanation $E(f, \mathcal{D})$.  As such, if the \emph{same} model $f$ and dataset $\mathcal{D}$ are used for all observations $i$ (as in the Forward Simulation use case introduced in the main text, then the global explanation $E(f, \mathcal{D})$ will be constant across all observations $i$.  We note that a researcher can choose to vary the predictive model $f_i$ and dataset $\mathcal{D}_i$ across observations $i$ (as in the Counterfactual Reasoning use case introduced in the main text; in this case the global explanation $E(f_i, \mathcal{D}_i)$ will vary across observations.
    \item \textbf{Local Problem + Local Explanation: } Like the previous example, the predictive base model can vary or be held constant across different observations $i$.  Each observation contains the base model's explanation on sampled point $\vx_i$.
\end{itemize}

\subsection{Encoding explanations in our experiments} \label{appdx:priorexps}

The explanations we consider in our experiment are LIME, SHAP, Anchors, and GAM: 
\begin{itemize}
    \item LIME and SHAP are both local explanations that return an importance score for each feature. We encode each explanation as a vector of length $d$ if $\vx \in \mathbb{R}^d$ and concatenate the explanation to the data-point to create an observation vector of length $2d$. Note that in Forward Simulation, to recreate the original setting, we extend the vector of length $d$ with additional information from the LIME package as discussed in Appendix \ref{sec:fwd_sim_results}.
    \item To get the GAM explanation, we used method \texttt{explain\_local} out-of-the-box from the InterpretML package \citep{nori2019interpretml}. 
    \item To get the Anchors explanation, we call the Anchors implementation~\citep{ribeiro2018anchors} which returns a set of anchors for a given data-point. For categorical features, the anchor is one of the values that the feature can take on. For continuous features, the anchor can take the form of a lower and upper bound. We include both the upper and lower bound for each continuous feature in the observation.  For a data-point $\vx$ with $d_{cat}$ categorical and $d_{cont}$ continuous features, the final observation has dimension at most $2 d_{cat} + 3 d_{cont}$. 
\end{itemize}

\subsection{Encoding other kinds of explanations}

There are other types of explanations that we do not consider in this work, but could also be studied using our proposed framework. We discuss some potential ways to encode these other types of explanations:

\begin{itemize}
    \item Saliency maps: These types of explanations provide a ``heat map" over an image input, where larger values signify a larger contribution of that pixel to the model's prediction. Suppose the dimensions of the input image are $d \times d \times c$ and the dimensions of the saliency map are $d \times d \times k$, one could define the observation as a $d \times d \times c+k$ input and instantiate the agent as a convolutional network. In Appendix~\ref{appdx:image}, we use this set up to replicate results for an image user study. Note that these types of explanations will most likely require a neural network-based agent architecture.
    \item Counterfactual explanations: These types of explanations provide the closest possible point that achieves the desired outcome for a given point. Suppose the data-points come from $\mathbb{R}^d$, then one possible encoding of the input observation is concatenating the two points into a $2d$ vector.
    \item Decision Sets: Suppose we are trying to encode an interpretable decision set~\citep{lakkaraju2016interpretable}, where each rule is written in if-else form and the data-points come from $\mathbb{R}^d$. One possible encoding of each rule is via a $2d$ vector where each feature has 2 slots (one for the equivalence operator and one for the value) and the slots would be filled in accordingly if the feature appeared in the rule. This way, if the interpretable decision set consists of $k$ rules, then the final encoding would be of the size $k \times 3d$. 
\end{itemize}

\section{Experimental Details}\label{appdx:expdetails}


\subsection{Dataset preprocessing} \label{preprocessing}

For forward simulation and data bugs, we used the UCI Adult (``adult'') dataset as done in~\citep{interpretinginterp, hase_evaluating_2020} (which is linked: \href{https://archive.ics.uci.edu/ml/datasets/Adult}{here}~\citep{Dua:2019}). The data was extracted by Barry Becker from the 1994 Census database. The citizens of the United States consent to having their data included in the Census. According to the U.S. Census Bureau, the United States/Commerce grants users a royalty-free, nonexclusive license to use, copy, and create derivative works of the Software. In the adult dataset, we dropped the \texttt{finalweight} feature. 
For all datasets, we one-hot encoded the categorical features and MinMax scaled all features. 

For counterfactual reasoning, we construct a $2$D synthetic regression dataset inspired by Kumar et al.~\citep{kumar2020problems}. 
We define the use case label as $1$ if increasing a data-point's specific feature $\vx_2$ increases the regression model prediction $f(\vx)$ (holding the other feature $\vx_1$ constant), $0$ otherwise. The data generation and other use case details are described extensively in Appendix~\ref{appdx:cfoverview}.
In the baseline setting, the agent/human receives no explanation. We evaluate the same local explanations as previous use cases (Section \ref{experiments}). 
We use a Light-GBM Regressor \citep{ke2017lightgbm} as the base prediction model class for all explanation settings except GAM, in which the GAM is the base model.

\subsection{Agent architecture} \label{deepsetarch}

We use the same Deepset architecture \citep{deepsets} for all three use cases (though, note that when the input is of set size 1 in the forward simulation and counterfactual reasoning use cases, Deepsets is effectively a standard feedfoward network). The Deepset architecture is characterized by two parts: (1) the feature extractor $\phi$, which encodes each item in the set, and (2) the standard network $\rho$, which takes in a summed representation of all of the representations from $\phi$. The feature extractor $\phi$ that we use has a sequence of Dense layers of sizes ($m$, where $m$ is the shape of each observation; 200; 100) respectively, where each layer is followed by an ELU activation with the exception of the last layer. The $\rho$ network is also a sequence of Dense layers of size (100, 30, 30, 10), each followed by an ELU activation with the exception of the last layer which is followed by a Sigmoid activation. 


\subsection{Missing Values Adjustment}\label{missingvalues}

\begin{figure}[h!]
\centering
  \includegraphics[scale=0.4]{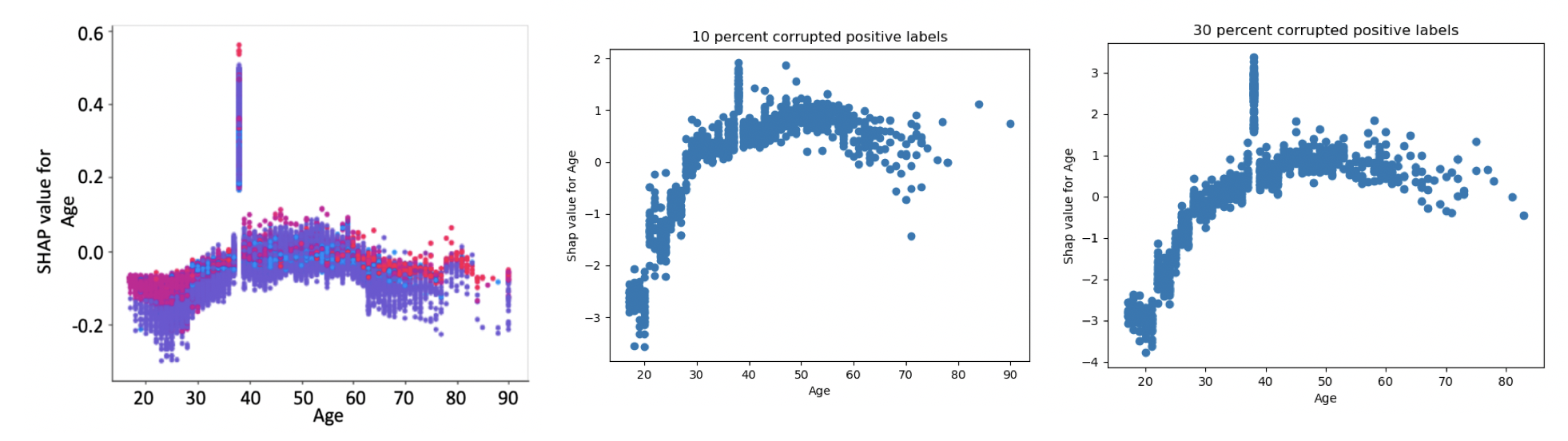}
  \caption{(Left) Image taken from the user interface from Kaur et al.~\citep{interpretinginterp} that plots the distribution of SHAP values for the age feature (Middle/Right) Distribution of SHAP values for 1000 points for 10\% and 30\% corruption.}
  \label{fig:missingvaluesshap}
\end{figure}

\begin{figure}[h!]
\centering
  \includegraphics[scale=0.6]{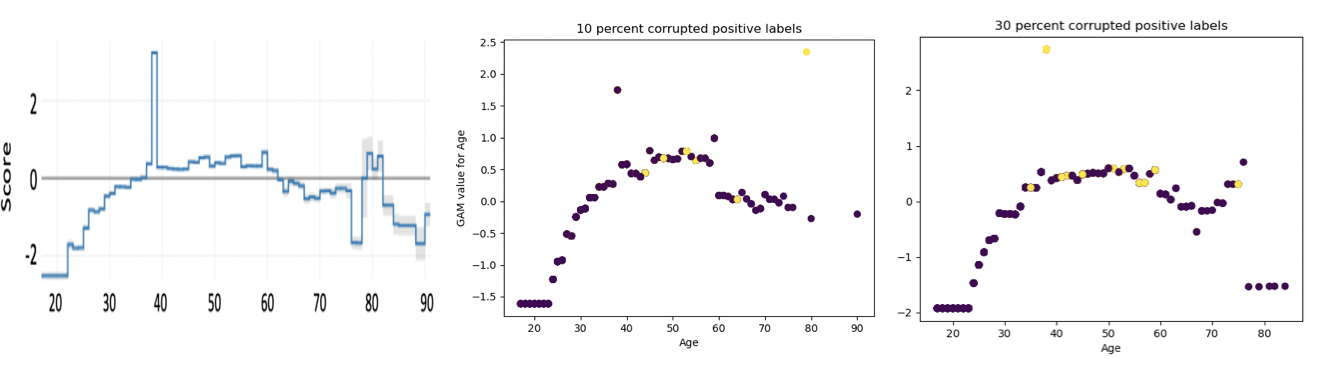}
  \caption{(Left) Image taken from the user interface from Kaur et al.~\citep{interpretinginterp} that plots the distribution of GAM values for the age feature (Middle/Right) Distribution of GAM values for 1000 points for 10\% and 30\% corruption.}
  \label{fig:missingvaluesgams}
\end{figure}

Figure \ref{fig:missingvaluesshap} and Figure \ref{fig:missingvaluesgams} show our attempt to recreate the original bug~\citep{interpretinginterp} as closely as possible.  The authors state that they apply the bug to 10\% of data-points.  However, we observe that the 30\% bug for both explanation methods much more closely resembles the figure that was presented in the original paper.

\subsection{Base Models}\label{sec:base-models}

Following Kaur et al.~\citep{interpretinginterp}, we evaluate both glass-box models that are designed to be interpretable (e.g., GAMs) and post-hoc explanation methods for black-box models (e.g., LIME, SHAP, Anchors).  Since GAM is glass-box, the GAM from which local explanations are derived \emph{is} the base predictive model.  On the other hand, LIME, SHAP, and Anchors are model-agnostic post-hoc explanation techniques, so they can be used to explain any predictive base model class that is compatible with their Python package implementations. This means that we need to additionally specify the base model family for post-hoc explanation techniques.

Ideally, we would have used the same underlying base predictive model for all explanation settings.  When the underlying base model is controlled and held constant, then differences in the agent's accuracy across explanation settings can be attributed to differences in the explanation methods - the variable we're interested in studying - instead of differences in the underlying base models. 

However, we encountered the same issue as Kaur et al.~\citep{interpretinginterp}: the InterpretML GAM implementation is not supported by the SHAP Python library; thus we could not generate SHAP explanations for a GAM predictive model.  As a result, we also chose to use a different base predictive model class for the 3 post-hoc explanation settings.  We emphasize that the comparisons that we make between explanation settings are particular to the base predictive model classes chosen: e.g. between the InterpretML implementation of a GAM and the SHAP Python package used to explain a LightGBM model.

\subsection{Framework Hyperparameters} \label{sec:hyperparams}

For all use cases, we generate an observation dataset of up to size 1000 and evaluate on 250 new observations. In our experiments we train our agent with the Adam optimizer with learning rate $10^{-4}$ for our optimizer (and weight decay $10^{-4}$ for counterfactual reasoning), weight decay of binary cross-entropy for the loss function, and batch size of $16$. We train for 350 epochs for forward simulation and data bugs and 1000 epochs for counterfactual reasoning. 



\subsection{Computational Resources}\label{appdx:compute}

In general, training an agent using the hyperparameters specified above does not require a GPU, making \simevals{} on particularly tabular datasets very accessible. When scaling up the data complexity, for example considering image-based data, it may be necessary to use GPUs to train both the prediction model and the agent model.

\section{Full results} \label{appdx:full_results}

We include a more complete set of experimental results for all three use cases. Average agent validation set accuracies (percentages) and their standard error are reported over 3 train-test splits.

\subsection{Forward Simulation} \label{sec:fwd_sim_results}

In Hase and Bansal~\citep{hase_evaluating_2020}, we noticed that the LIME explanation provided included (a) the LIME approximation model's weights, (b) the approximation model's intercept, (c) the sum of the approximation model's weights, and (d) the approximation model's prediction on the point. We also include results for a setting where the agent is given the LIME approximation model's weights only.  While the agent's accuracy in this setting is lower, it still outperforms the baseline setting. 

\begin{table}[H]
\centering
\caption{Error bars when varying the training observations the agent receives to perform forward simulation. \\}
\begin{tabular}{@{}llllll@{}}
\toprule
                 & \multicolumn{4}{c}{Number of Train Set Observations} \\
Explanation & 16 & 32 & 100 & 1000\\ \midrule
LIME     & 94.2\% $\pm$  3.3\%  & 99.8\% $\pm$  0.2\%  &   100.0\% $\pm$  0.0\%    &     100.0\% $\pm$  0.0\%   \\
LIME (Weight only)      & 86.3\% $\pm$  5.5\%  & 90.8\% $\pm$  1.7\%  &   91.8\% $\pm$  1.2\%    &     96.3\% $\pm$  0.2\%   \\
Anchors   & 89.2\% $\pm$ 2.0\%  & 93.5\% $\pm$  3.4\%  &   94.7\% $\pm$  2.5\%    &     93.7\% $\pm$  1.5\%  \\
SHAP       & 94.5\% $\pm$  2.6\%  & 97.3\% $\pm$  0.3\%  &   99.1\% $\pm$  0.6\%    &     99.3\% $\pm$  0.5\%     \\
GAM        & 89.5\% $\pm$  9.6\%  & 96.3\% $\pm$  2.3\%  &   97.4\% $\pm$  0.2\%    &     98.8\% $\pm$  0.3\%    \\
No explanation   & 82.3\% $\pm$  1.3\%  & 83.7\% $\pm$  1.5\%  &   85.7\% $\pm$  1.8\%    &     88.7\% $\pm$  0.4\%   \\ \bottomrule
\end{tabular}
\label{appd:forwardsim}
\end{table}

We note that for forward simulation, there exist known heuristics to predict the model's output using some explanation methods (i.e., it is possible for a human or agent to achieve (near) perfect accuracy). For example, SHAP feature attributions by definition sum to the model prediction \citep{lundberg2017unified}. With sufficient examples ($N > 1000$), the predictiveness of SHAP explanations for this use case is demonstrated by the agent's near perfect accuracy.

For completeness we provide the $p$-values from Hase and Bansal~\citep{hase_evaluating_2020} which are: LIME (0.014) and Anchors (0.234). Their statistical analysis only compares each explanation to the baseline of no explanation, so the $p$-values provided signify whether the difference in test accuracy of the human when given an explanation versus not is statistically significant.

\subsection{Counterfactual Reasoning}\label{appdx:cfresults}

We provide results for the counterfactual reasoning use case for \simevals{} as well as our MTurk study. We find that \simevals{} performs well using the LIME explanation which is what we observe in human performance as well.

\begin{table}[H]
\centering
\caption{We vary the number of training observations the agent receives to perform counterfactual reasoning.\\}
\begin{tabular}{@{}llllll@{}}
\toprule
 & \multicolumn{5}{c}{Number of Train Set Observations}   \\ 
Explanation & 4 & 16 & 64 & 100 & 1000   \\ \midrule
LIME        & 92.9\% $\pm$ 1.7\% & 94.5\% $\pm$ 2.9\% & 98.8\% $\pm$ 0.1\% & 99.2\% $\pm$ 0.8\% & 99.7\% $\pm$ 0.1\%   \\
SHAP         & 56.3\% $\pm$ 2.3\% & 52.5\% $\pm$ 1.3\% & 55.3\% $\pm$ 1.8\% & 57.3\% $\pm$ 2.5\% & 64.9\% $\pm$ 3.1\%     \\
GAM         &  56.0\% $\pm$ 2.1\% & 53.5\% $\pm$ 1.8\% & 56.1\% $\pm$ 2.9\% & 57.3\% $\pm$ 3.2\% & 63.5\% $\pm$ 1.8\%   \\
Model Prediction    & 52.1\% $\pm$ 1.6\% & 55.6\% $\pm$ 1.7\% & 54.7\% $\pm$ 2.7\% & 57.9\% $\pm$ 2.4\% & 60.3\% $\pm$ 2.9\%    \\ \bottomrule
\end{tabular}
\label{table:cfreasoning}
\end{table}

\begin{table}[h!]
\centering
\caption{The average user accuracy on 15 train and test observation along with standard error for counterfactual reasoning use case where for each explanation setting we recruited $N=20$ Turkers.}
\begin{tabular}{@{}lll@{}}
\toprule
Explanation    & \multicolumn{1}{c}{Train}   & \multicolumn{1}{c}{Test}    \\ \midrule
LIME           & 66.6 $\pm$ 23.2\% & 69.4 $\pm$ 28.2\% \\
SHAP           & 50.3 $\pm$ 8.9\% & 41.4 $\pm$ 14.0\% \\
GAM            & 48.7 $\pm$ 12.1\% & 45.7 $\pm$ 16.4\%\\
No Explanation & 58.4 $\pm$ 15.4\% & 48.6 $\pm$ 12.8\% \\ \bottomrule
\end{tabular}
\label{tab:mturkresults}
\end{table}

\begin{table}[H]
\centering
\caption{Pairwise comparisons using Tukey's HSD between explanation conditions for the counterfactual reasoning MTurk study, showing p-values. We consider $p < 0.05$ to be statistically significant.}
\begin{tabular}{|l|l|l|l|}
\hline
       & LIME            & SHAP  & GAM   \\ \hline
LIME   &                 &       &       \\ \hline
SHAP   & \textbf{0.0017} &       &       \\ \hline
GAM    & \textbf{0.0016} & 0.906 &       \\ \hline
No Exp & \textbf{0.007}  & 0.996 & 0.665 \\ \hline
\end{tabular}
\end{table}

\subsection{Missing Values}

We provide results for the missing values data bug use case for \simevals{} as well as our MTurk study. We find that \simevals{} performs well using the SHAP and GAM explanation which is what we observe in human performance as well. 

\begin{table}[H]
\centering
\caption{Recreating Kaur et al.~\citep{interpretinginterp}'s missing values setting, showing that both SHAP and GAM were successful in finding the bug with high accuracy. 
We additionally vary the size of the observation set $S$.\\}
\begin{tabular}{@{}lllll@{}}
\toprule
                 & \multicolumn{4}{c}{Observation Set Size}   \\ 
Explanation      & 1      & 10     & 100    & 1000      \\ \midrule
SHAP             & 63.2\% $\pm$  2.4\%  & 84.0\% $\pm$  1.2\%  &   99.8\% $\pm$  0.2\%    &     100.0\% $\pm$  0.0\%     \\
GAM             & 64.8\% $\pm$  3.1\%  & 87.7\% $\pm$  3.1\%  &   100.0\% $\pm$  0.0\%    &     100.0\% $\pm$  0.0\%    \\
LIME             & 55.2\% $\pm$  1.4\%  & 56.9\% $\pm$  1.3\%  &   64.5\% $\pm$  1.5\%    &     75.9\% $\pm$  0.8\%    \\
Model Prediction  & 57.5\% $\pm$  1.4\%  & 57.4\% $\pm$  1.2\%  &   58.2\% $\pm$  2.0\%    &     67.3\% $\pm$  11.2\%   \\ \bottomrule
\end{tabular}
\label{appd:missingvaluesvary1}
\end{table}

\begin{table}[H]
\centering
\caption{Standard errors when varying the strength of the missing values bug for a fixed observation set size of $1000$.\\}
\begin{tabular}{@{}lllll@{}}
\toprule
                 & \multicolumn{4}{c}{Bug Strength}   \\ 

Explanation      & 5\%    & 10\%   & 20\%   & 30\%   \\ \midrule
SHAP              & 81.25\% $\pm$  3.6\%  &   82.3\% $\pm$  28.0\%    &     100.0\% $\pm$  0.0\%   &   100.0\% $\pm$  0.0\%    \\
GAM              & 58.7\% $\pm$  0.8\%  &   75.2\% $\pm$  21.6\%    &     100.0\% $\pm$  0.0\%   &   100.0\% $\pm$  0.0\%    \\
LIME             & 60.9\% $\pm$  2.4\%  &   61.8\% $\pm$  0.4\%    &     60.2\% $\pm$  4.7\%   &   68.7\% $\pm$  11.9\%    \\
Model Prediction  & 54.4\% $\pm$  2.0\%  &   56.1\% $\pm$  2.6\%    &     56.6\% $\pm$  2.9\%   &   65.0\% $\pm$  12.4\%    \\ \bottomrule
\end{tabular}
\label{missingvaluestrength1}
\end{table}

\begin{table}[h!]
\centering
\caption{The average user accuracy on 15 train and test observation along with standard error for data bugs (missing values) use case where for each explanation setting we recruited $N=20$ Turkers. }
\begin{tabular}{@{}lll@{}}
\toprule
Explanation    & \multicolumn{1}{c}{Train}   & \multicolumn{1}{c}{Test}    \\ \midrule
SHAP           & 60.7\% $\pm$ 19.4\% & 67.4\% $\pm$ 27.1\% \\
GAM           & 58.1\% $\pm$ 16.2\% & 64.4 $\pm$ 15.6\% \\
LIME            & 53.3\% $\pm$ 11.2\% & 48.0\% $\pm$ 12.3\%\\
Model Prediction & 46.7\% $\pm$ 13.7\% & 40.7\% $\pm$ 11.9\% \\ \bottomrule
\end{tabular}
\label{tab:mturkresults}
\end{table}

\begin{table}[h!]
\centering
\caption{Pairwise comparisons between explanation conditions for the missing values (data bugs) MTurk study, showing p-values. We consider $p < 0.05$ to be statistically significant.}
\begin{tabular}{|l|l|l|l|}
\hline
                                                          & LIME           & \begin{tabular}[c]{@{}l@{}}Model Prediction\end{tabular} & SHAP  \\ \hline
LIME                                                       &                &                                                            &       \\ \hline
\begin{tabular}[c]{@{}l@{}}Model Prediction\end{tabular} & 0.615          &                                                            &       \\ \hline
SHAP                                                       & \textbf{0.041} & \textbf{0.001}                                             &       \\ \hline
GAM                                                        & \textbf{0.036} & \textbf{0.001}                                             & 0.998 \\ \hline
\end{tabular}
\end{table}

\subsection{Redundant Features}\label{appdx:redfeat}

To create a setting where there is no bug, we randomize the value of one of the features so there is no correlation between the two and only one feature contains the original information. Note that, in the original user study, the authors do not vary the \emph{strength} of the bug, but we are easily able to study this variant of the bug with our algorithmic framework. We increase the strength of the bug by increasing the number of data-points in dataset $\mathcal{D}_i$ that have the two correlated features. 

The agent's accuracy was near random guessing when given LIME explanations or the model prediction baseline, suggesting that neither would be helpful for a human user.

\begin{table}[H]
\centering
\caption{We vary the strength of the redundant features bug on the Adult dataset for a fixed observation set size of 1000 and corroborate results from Kaur et al.~\citep{interpretinginterp}.\\}
\begin{tabular}{@{}llllll@{}}
\toprule
 & \multicolumn{5}{c}{Bug Strength}   \\ 
Explanation & 10\% & 30\%   & 50\% & 70\%   & 90\%   \\ \midrule
SHAP         & 74.7\% $\pm$  2.5\%  & 78.3\% $\pm$  2.6\%  &   87.5\% $\pm$  3.7\%    &     96.8\% $\pm$  2.5\%   &   99.7\% $\pm$  0.2\%    \\
GAM         & 56.5\% $\pm$  1.2\%  & 58.9\% $\pm$  2.3\%  &   84.0\% $\pm$  26.9\%    &     69.8\% $\pm$  25.7\%   &   71.9\% $\pm$  24.7\%    \\
LIME        & 56.7\% $\pm$  1.3\%  & 57.3\% $\pm$  2.5\%  &   54.5\% $\pm$  2.6\%    &     54.8\% $\pm$  3.9\%   &   55.3\% $\pm$  1.3\%    \\
Model Prediction     & 57.6\% $\pm$  3.0\%  & 56.6\% $\pm$  2.0\%  &   53.7\% $\pm$  2.9\%    &     55.2\% $\pm$  0.8\%   &   58.8\% $\pm$  2.7\%    \\ \bottomrule
\end{tabular}
\label{table:redfeat}
\end{table}

\subsection{Ablation results}\label{appdx:ablation}

\textbf{Single data-point observation:} Counterfactual reasoning is one of the use cases where the agent is presented with a single data-point (and explanation). In the main text, we presented results using the DeepSet architecture. Here, we swap out DeepSet with a non-neural network architecture, LightGBM. We find that the rankings of the explanations are consistent (e.g., compare Table~\ref{table:cfreasoning} and Table~\ref{table:cfreasoning1}).

\begin{table}[H]
\centering
\caption{Counterfactual reasoning ablation: we switch out the DeepSet architecture for LightGBM. We find that the rankings are still consistent with the DeepSet results.\\}
\begin{tabular}{@{}llllll@{}}
\toprule
 & \multicolumn{5}{c}{Number of Train Set Observations}   \\ 
Explanation & 4 & 16 & 64 & 100 & 1000   \\ \midrule
SHAP         & 48\% & 48\% & 47\%& 52\% & 58\%     \\
GAM         &  48\% & 48\% & 49\%& 50\% & 57\%   \\
LIME        & 48\% & 48\% & 99.6\%& 99.6\% & 99\%    \\
Model Prediction    & 48\% & 48\% & 48\%& 53\% & 52\%    \\ \bottomrule
\end{tabular}
\label{table:cfreasoning1}
\end{table}

\textbf{Multiple data-point observation.} Data bugs is a use case where the user receives multiple data-points as the observation. We do not include these results, but we find low accuracy when using non-neural network models like LightGBM. However, we find similar results when ablating the neural network architecture. For example, instead of summing the representations (as is done in DeepSet), we instead concatenate representations. We find comparable results (with the caveat that one might need to scale up the number of training points in the dataset). For example, for SHAP with set size 100, one would need 4.5x larger dataset to achieve the same accuracy as the original Deepset model (99.8\%): 1x data (57.9\%), 1.5x data (78.9\%), 2.5x data (92.3\%), 3.5x data (95.6\%), 4.5x data (99.5\%). This is because concatenating representations drastically increases the model parameter size, particularly for larger set sizes. We also find that the agent architecture is not drastically affected by adding/removing a layer (Table~\ref{appd:missingvaluesvary1} vs Table~\ref{missingvaluesvary})



\begin{table}[h!]
\centering
\caption{Adding a layer to DeepSet, for bug strength of 30\%\\}
\begin{tabular}{@{}llll@{}}
\toprule
                 & \multicolumn{3}{c}{Observation Set Size} \\
Explanation            & 10       & 100      & 1000    \\ \midrule
SHAP              & 85.8\%   & 99.5\%   & 100\%  \\
GAM            & 90.0\%   & 100\%   & 100\%  \\
LIME                & 53.3\%   & 65.8\%   & 73.5\%  \\ \bottomrule
\end{tabular}
\label{missingvaluesvary_ablation}
\end{table}

\section{MTurk Details} \label{appdx:mturk}



\subsection{Participants.}

A total of 80 participants were recruited to complete the task via Amazon MTurk in 4 batches (20 for each explanation setting).  We controlled the quality of the Turkers by limiting participation in our MTurk HIT adults located in an English-speaking countries (specifically the United States and Canada) with greater than a 97\% HIT approval rate for quality control.  Each worker was only allowed to complete the study 1 time.  All participants were retained for final analysis.  The estimated time to complete the study was 15 minutes.  The study took 21 minutes on average (however, this includes any pauses/breaks the Turker might have taken between the two portions of the task).  Each worker was compensated 2.50 USD for an estimated hourly wage of 10 USD.

Our study was approved by the IRB (STUDY2021\_00000286) for exempt Category 3 research. We did not anticipate any potential participant risks as we described in the IRB:  ``The risks and discomfort associated with participating in this study are no greater than those ordinarily encountered in daily life or operation of a smartphone or laptop, such as boredom or fatigue due to the length of the questionnaires or discomfort with the software.'' 


\subsection{Study Context.}

\textbf{Counterfactual Reasoning.} We introduce the Turkers to the use case using the context of furniture pricing: Turkers are shown data about a furniture item (height and length measurements $\vx_i$ and the item's price $f(\vx_i)$) and are asked to perform counterfactual reasoning by choosing whether or not increasing the item's length would increase its price (a screenshot of the interface is shown in Figure \ref{fig:turk-exp-settings}).  The Turkers are instructed to set aside any prior knowledge they may have about furniture pricing and only use the information provided during the study.

\textbf{Data Bugs.} We introduce the Turkers to the use case using the context of identifying ``bugs'' or problems in an Artificial Intelligence system which is used to make predictions about individual incomes. Turkers are shown scatter plots that follow the presentation from the original user study conducted by Kaur et al.~\citep{interpretinginterp}, reflecting how the system is making predictions along each attribute. The user is provided a demo showing that, for example, as the amount of education a person receives increases, the system deems that attribute to be more important to income prediction. The participants are then asked to determine whether the system contains a ``bugs'' and learn from the feedback. A screenshot of the interface is shown in Figure \ref{fig:turk-instruct-databugs}.

\subsection{Study Interface.}

Participants were first presented with brief information about the study and an informed consent form.  To complete the HIT, participants were instructed to finish both parts (corresponding to a ``Train'' and ``Test'' phase) of the HIT via external URLs to Qualtrics surveys.  Both phases of the MTurk evaluation task described in the main text were conducted entirely on Qualtrics. 

In both phases of the study, participants are presented with a series of furniture items, and are instructed to answer survey questions where they predict if increasing the item's length will increase its price.  Figure \ref{fig:turk-exp-settings} illustrates how the information shown to the user varies across different explanation settings.

\begin{figure}[H]
\centering
  \includegraphics[scale=0.60]{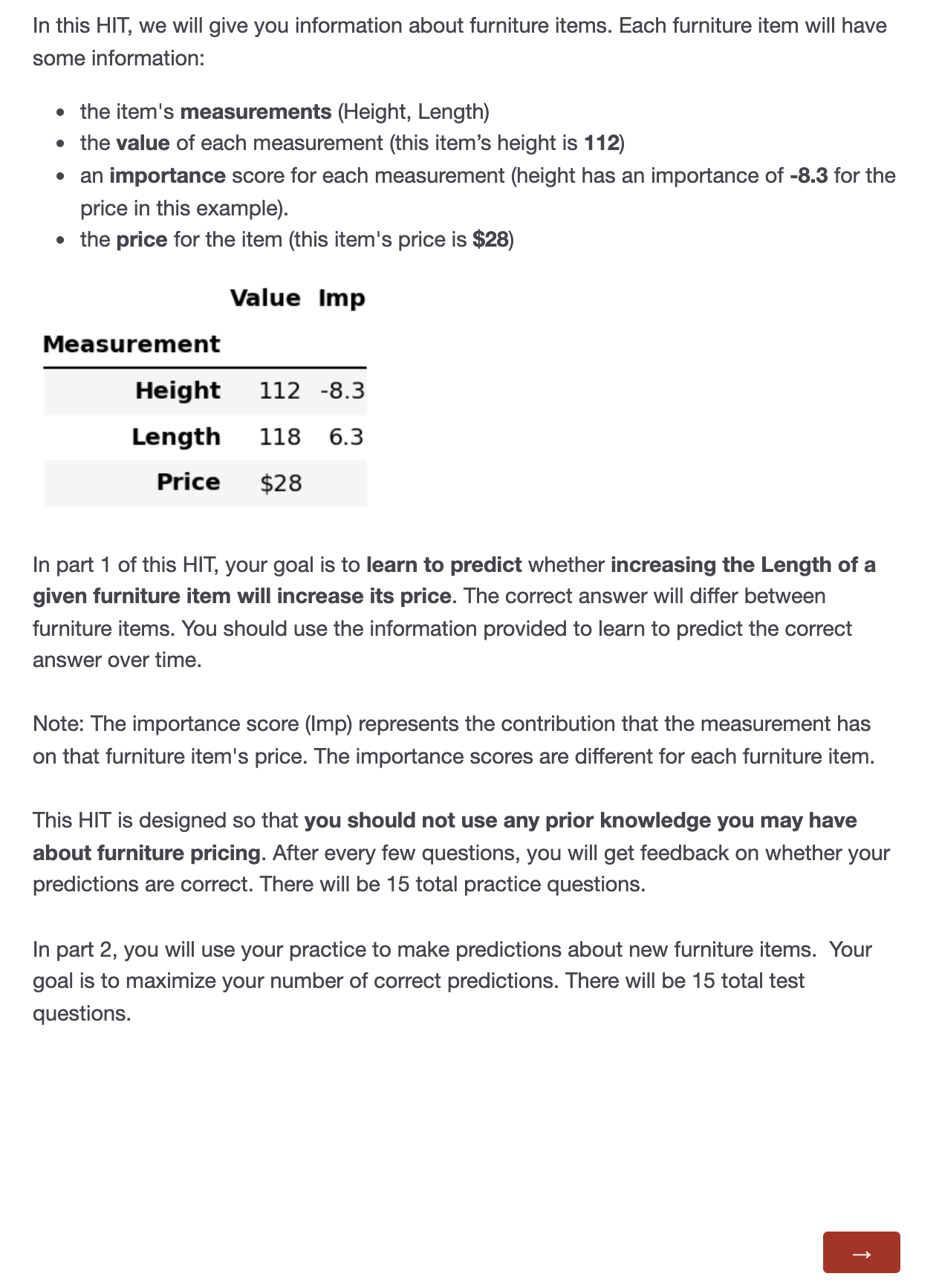}
  \caption{Instructions given to participants in the setting where \textbf{explanations are provided} upon opening the Qualtrics survey for the Train phase.  The instructions (a) introduce the furniture pricing task and terminology used in the study, (b) show and explain how to interpret an example observation, (c) instruct the Turkers to complete the task by learning from feedback given (and to not use their prior knowledge), and (d) describe the two-phase format of the study. Note that the instructions given to participants in the \textbf{baseline setting} where explanations are not provided is the same as the above except that the example observation \emph{does not} have an importance score column and the term ``importance score'' is also never defined. }
  \label{fig:turk-instructions}
\end{figure}

\begin{figure}[H]
\centering
  \includegraphics[scale=0.60]{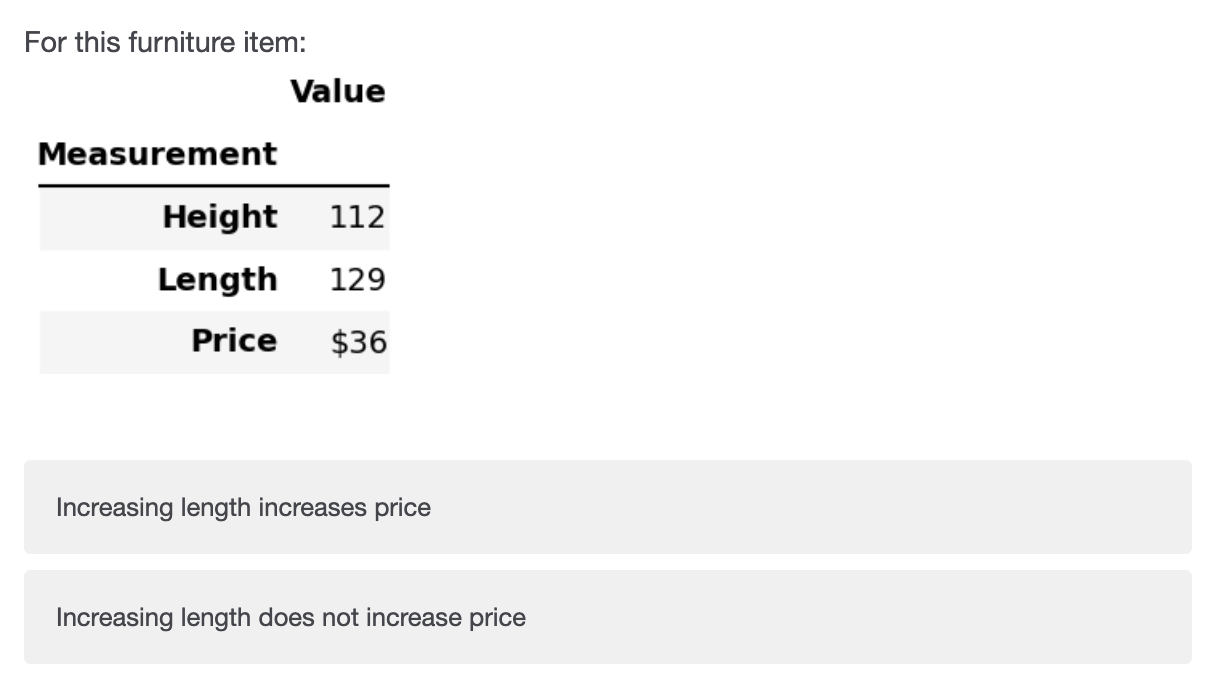}\\
  - - - - - - - - - - - - - - - - - - - - - - - - - - - - - - - - - - - - - - - - - - \\
  \includegraphics[scale=0.60]{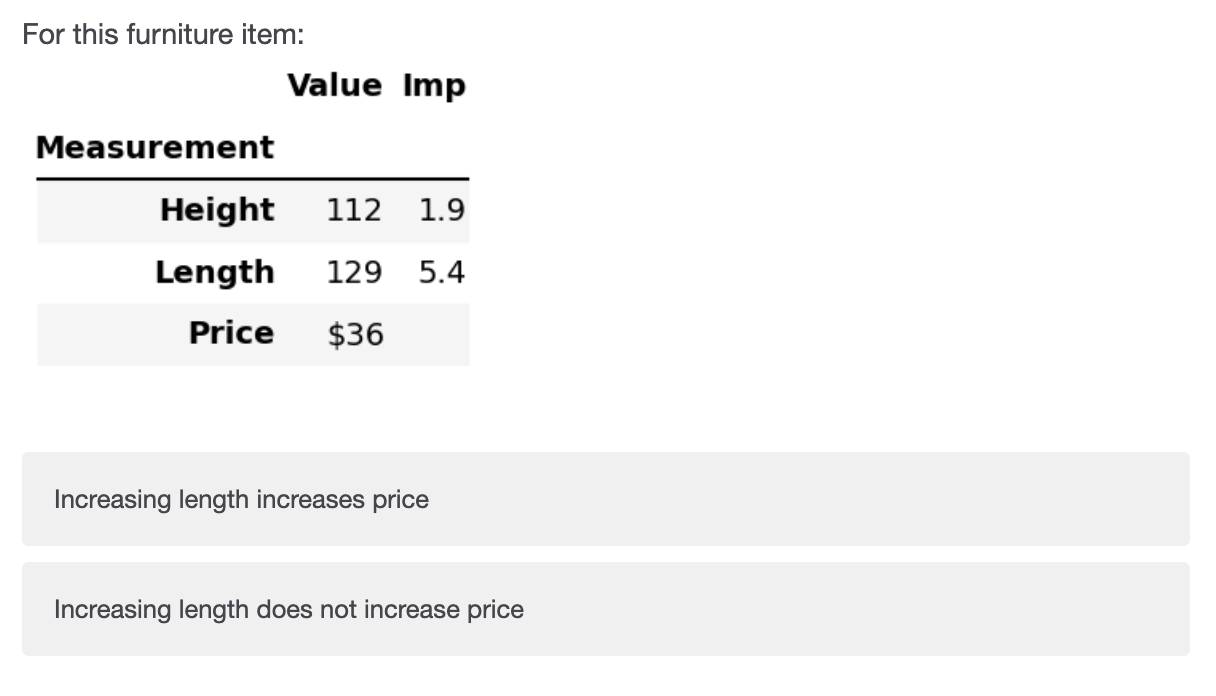}\\
  \caption{The information provided to participants varies for the Baseline, No Explanation (Top) and the LIME (Bottom) explanation settings.  For the LIME, SHAP, and GAM explanation settings, the participant is given the explanation's assigned importance score for each feature in the ``Imp'' column.  Participants are not provided with feature importance scores in the Baseline, No Explanation setting.  Each participant is assigned to 1 explanation setting for the entirety of the study, and so will only ever see 1 of these possible interfaces. }
  \label{fig:turk-exp-settings}
\end{figure}

\begin{figure}[H]
\centering
  \includegraphics[scale=0.47]{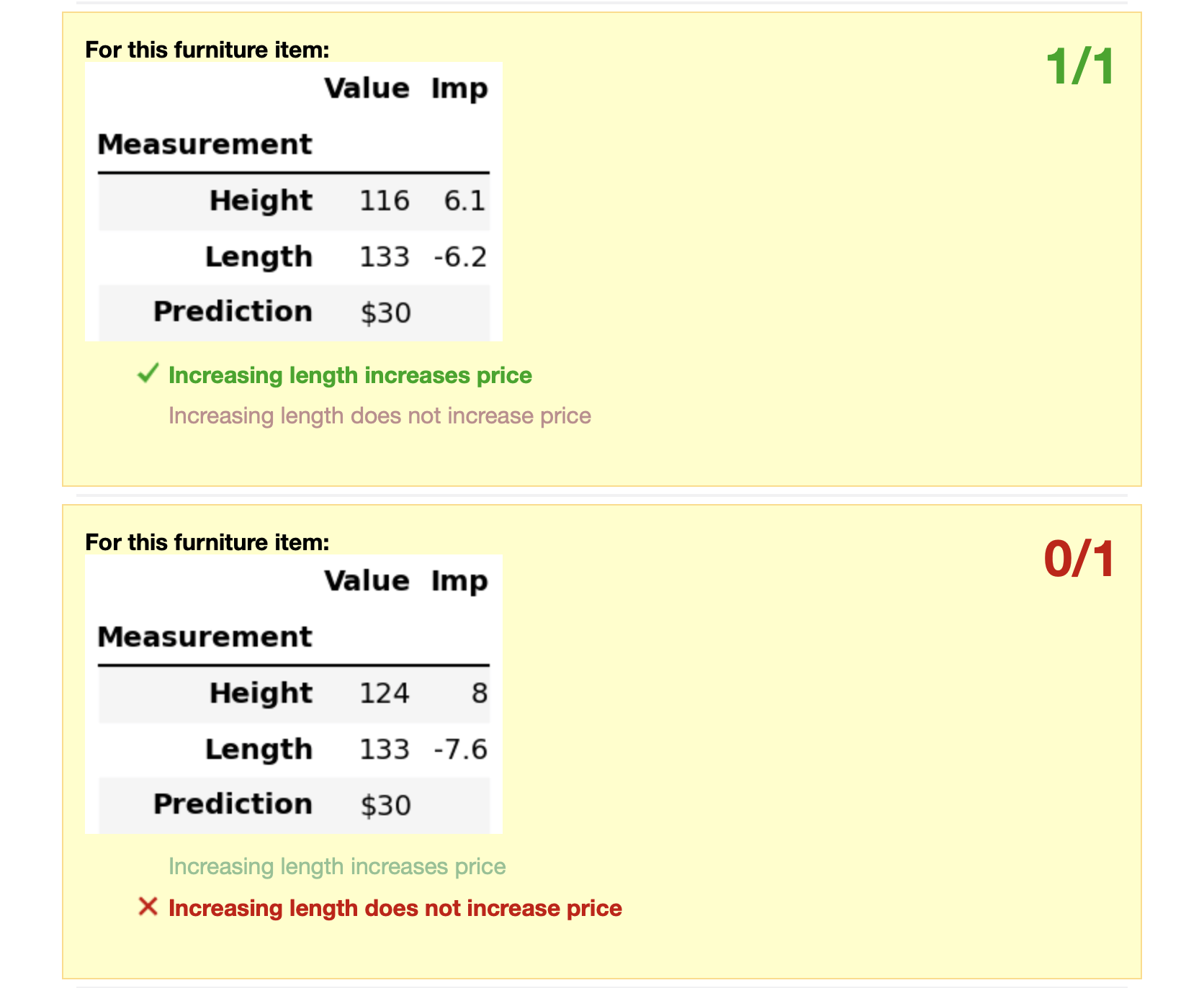}
  \caption{In the Train phase only, participants are provided with feedback on their responses for each furniture item.  After submitting their responses, participants learn if their response is correct (Top Example) or incorrect (Bottom Example). Participants may only submit 1 response per furniture item (e.g. they cannot modify their response after it has been submitted), and receive feedback after submitting responses for all 5 observations on each page.}
  \label{fig:turk-feedback}
\end{figure}

\begin{figure}[H]
\centering
  \includegraphics[scale=0.47]{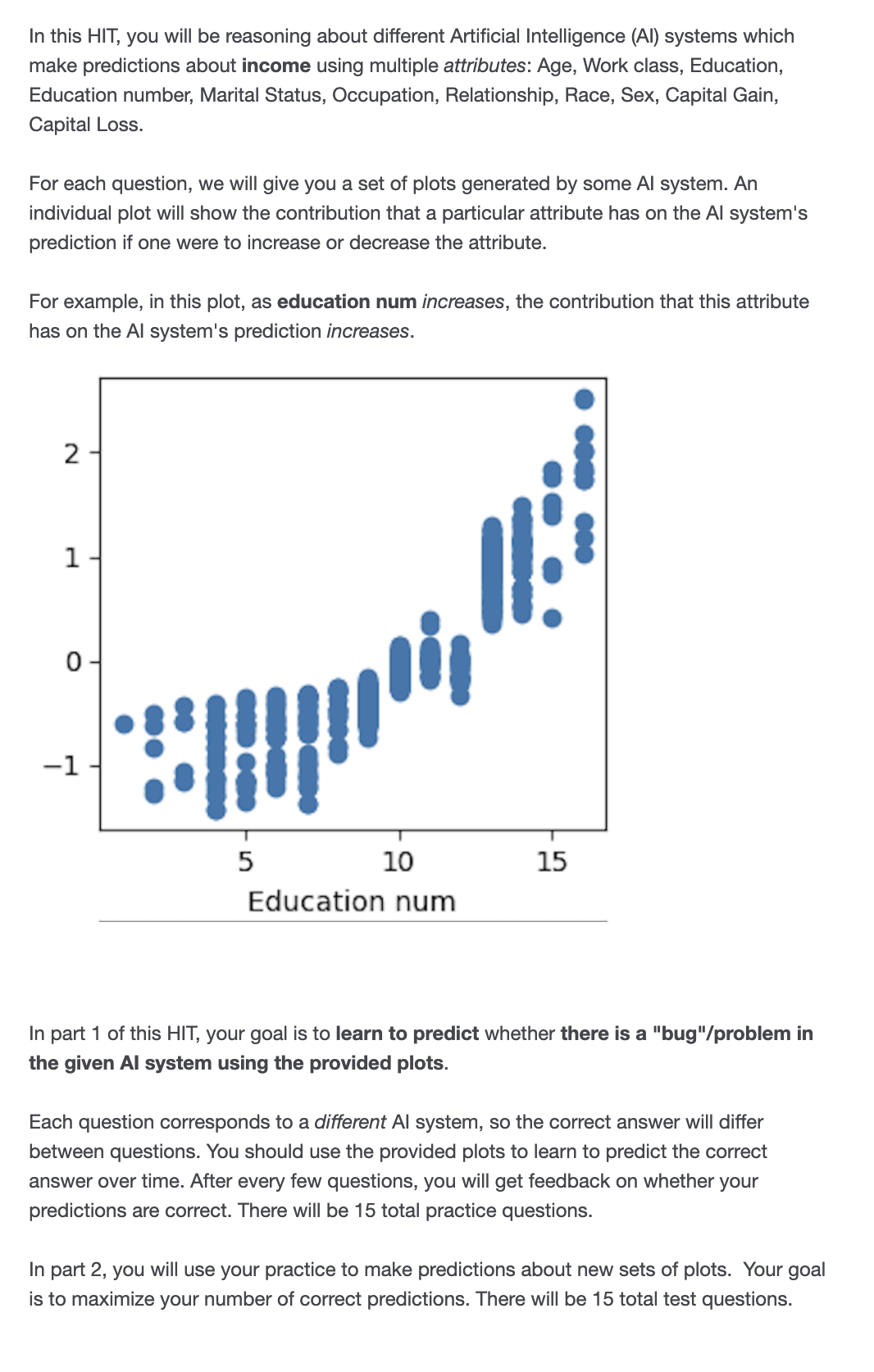}
  \caption{For the data bugs user study, we followed a similar introduction to the task as in the counterfactual reasoning set-up, but instead of a furniture task we tell the Turkers they are finding problems/bugs in an Artificial Intelligence system.}
  \label{fig:turk-instruct-databugs}
\end{figure}

After opening the Train survey, participants are presented with instructions about the counterfactual reasoning task (Figure \ref{fig:turk-instructions}).  After reading the instructions, participants begin the Train phase of the study, where they receive feedback after submitting their responses (Figure \ref{fig:turk-feedback}). Participants then complete the Test phase of the study, where they are not given any feedback after submitting their responses (but are presented with an otherwise identical interface to the Train phase).

For each question, we recorded the participants' response and the time the participant took to respond.  At the end of the Test phase, we asked participants to describe the strategy that they used to answer the questions. We also included an attention question where we asked the Turkers to list the two measurements provided. We found that all Turkers either fully answered this question correctly (by saying Height and Length) or gave a response that was reasonable (by saying Price and Length).


\begin{figure}[H]
\centering
  \includegraphics[scale=0.3]{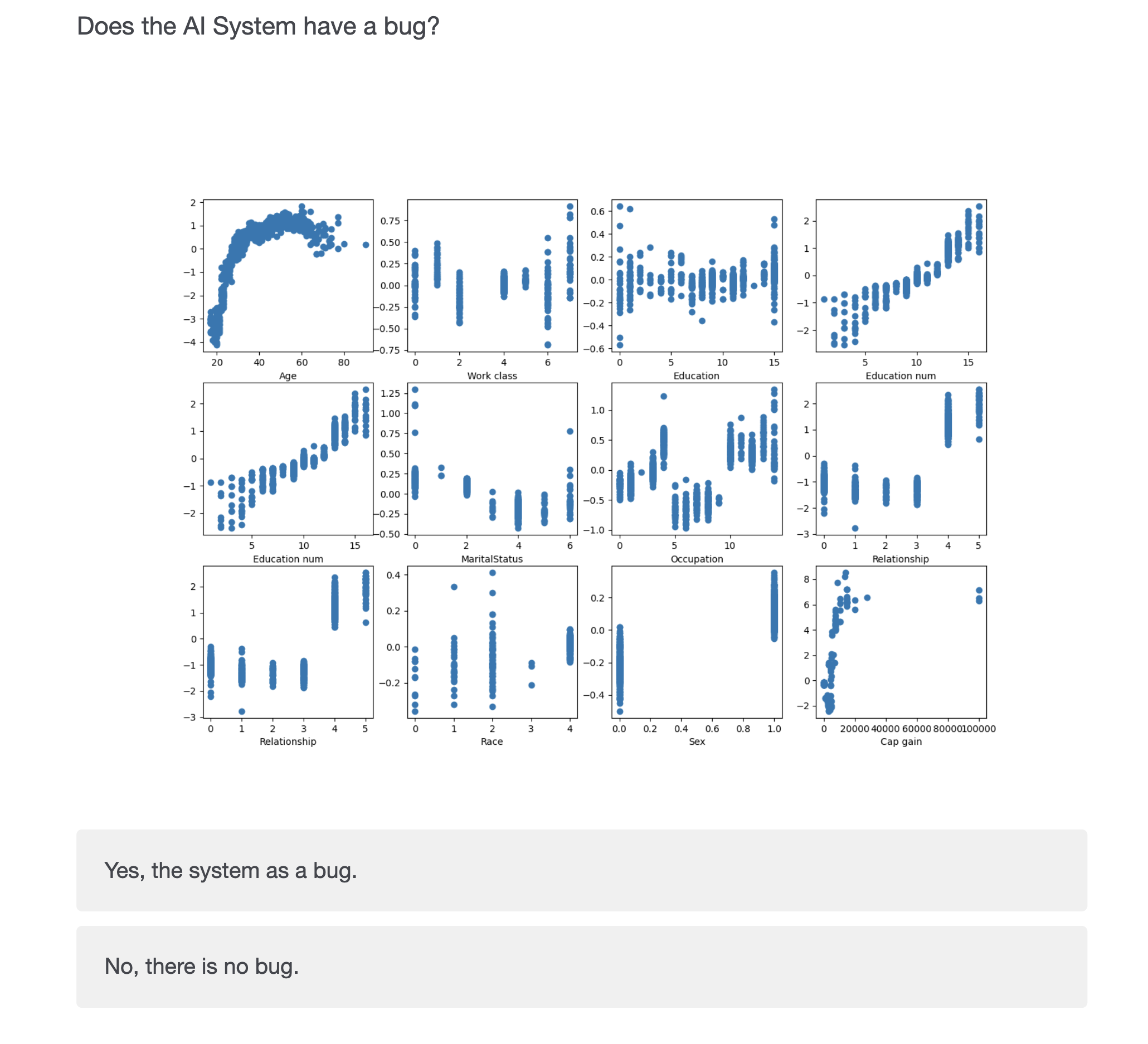}
  \caption{Following the visualizations presented in Kaur et al.~\cite{interpretinginterp}, we aggregated the explanations of set size $S=1000$ along each attribute. Specifically, we subset the explanation score for each attribute and plot it against the attribute's values. }
  \label{fig:turk-q-databugs}
\end{figure}

\subsection{Pilot Studies} \label{sec:pilot}

We conducted several pilot human subject studies that informed the design of our final MTurk study.  Our pilot studies showed that subjects struggled to complete the counterfactual reasoning task when presented with data $\vx$ with a 3 or more features, likely due to information overload~\citep{poursabzi2018manipulating}.  Thus we chose to construct a 2D dataset for our final study.

Our pilot studies also showed that instructions presented to subjects influenced the strategy that they used to respond to the questions.  Particularly, we noticed that the way we defined \emph{importance scores} influenced how participants used the scores when answering each question.  In an initial trial, we provided participants with the following more detailed definition of importance scores:

\begin{itemize}
    \item \texttt{A positive importance score means that the measurement’s value contributes to the furniture item having a high price.}
    \item \texttt{A negative importance score means that the measurement’s value contributes to the furniture item having a low price.}
    \item \texttt{An importance score with a large magnitude (absolute value) means that the measurement’s value has a larger contribution to determining the item’s price.}
\end{itemize}

However when we reviewed the participants' self-reported descriptions of the strategy they used, several participants reported that they ``used the strategy provided by the instructions'' rather than learn from the feedback given during the Train phase.  While the given instructions do not directly state a strategy for how the importance scores should be used for the task when literally interpreted, the instructions can be \emph{misinterpreted} to mean that a positive length importance score implies that the item's price will \emph{increase} when its length is increased.  Indeed, this is the strategy that many participants used when given the above instructions.  As such, we decided to provide a more minimal description of importance scores in the final study (shown in Figure \ref{fig:turk-instructions}) to encourage participants to learn their own strategy using the information provided as opposed to simply following the given instructions.

\section{Image Experiment}\label{appdx:image}

We also conducted \simevals{} for a user study which evaluated saliency maps for model debugging~\citep{debuggingtests}. We provide this example primarily to illustrate how to perform \simevals{} on saliency maps and chose not to include these experiments in the main text because there were no positive results for saliency maps (e.g., providing the model prediction alone was enough to detect the bug). The base dataset for this user study was a dog breed dataset (the image label is the breed of the dog). The authors conducted a user study on 4 different bugs: out-of-distribution input, label errors, spurious correlations, and top layer randomization. The authors find that for all of the use cases except for spurious correlations, users relied on \emph{the model prediction} to detect the bug as opposed to the \emph{explanation}. Unfortunately, due to the lack of available code, we were unable to recreate the same spurious correlations that were studied in the paper.

We trained \simevals{} on the label error bug. We followed a similar data generation process as in the Data Bugs use cases described in the main text and introduced label errors by randomizing the labels before the prediction model was trained and attempt to recreate the prediction model described in the paper. We instantiated the agent architecture using a convolutional Deepset architecture by modifying the $\phi$ network to be a series of convolutions.

The architecture for the agent that is provided with saliency maps: For both the saliency map and the original image, we pass each through a ResNet50 model as a feature extractor, flatten, dropout with probability 0.5, Linear layer with 128 nodes and ReLU non-linearity. We concatenate the two vectors and then pass through another Linear layer with 100 nodes and ReLU non-linearity, and finally a Linear layer with 1 output and Sigmoid non-linearity.

The architecture for the agent that is provided with only the model prediction: For the original image, we pass each through a ResNet50 model as a feature extractor, flatten, dropout with probability 0.5, Linear layer with 128 nodes and ReLU non-linearity. The model prediction is encoded as a one-hot vector and is passed through a Linear layer with 100 nodes and ReLU non-linearity. We concatenate the two vectors and then pass through another Linear layer with 100 nodes and ReLU non-linearity, and finally a Linear layer with 1 output and Sigmoid non-linearity.

We use batch size of 16, Adam optimizer with learning rate 0.001, binary cross entropy loss, and train for 10 epochs.

\textit{Findings:} We find that the agent is able to achieve $>95\%$ accuracy when provided with the model prediction. This makes sense because provided that the model has a reasonable mental model of classifying which breed a dog is, then it is easy to tell whether the image is correctly classified or not. However, when provided with saliency maps (e.g., Gradient, Integrated Gradient), the agent accuracy is $~50-65\%$, despite attempts to ablate the CNN architecture. After visual inspection (similar to Figure 13 of their paper) of the saliency map for both cases, it is not particularly evident why it would be helpful in distinguishing when there is a label error.



\section{Discussion: The Agent-Human Gap}\label{appdx:humanfactors}


\subsection{Human Factors}

We describe several human factors that should be considered when researchers are designing \simevals{} and interpreting a \simeval{} agent's accuracy. We also discuss how we considered these factors when designing our own MTurk study.

\textbf{Proxy Metrics.} Since \simevals{} is agnostic to whether a provided explanation is faithful or not to the model that it is explaining.  A researcher using \simevals{} to select candidate explanation methods should be careful to check that the explanations under consideration also satisfy desirable proxy metrics. This way one would not select explanations that may inadvertently mislead the users. Important proxy metrics include faithfulness to the model \citep{ribeiro2016should, plumb2018model} and stability over runs \citep{alvarez, agarwal2022rethinking, ghorbani2019interpretation}. In our user study, we computed explanations using well known, open-source explanation methods and ensured not to ``adversarially'' select explanations which would mislead the humans but enable the agent to achieve high accuracy.

\textbf{Complexity of Explanation.} A well-studied human factor is that more complex explanations will negatively impact a human's ability to use the explanation~\citep{poursabzi2018manipulating,lage2019evaluation}. Provided with enough data, the algorithmic agent may in some cases perform better with more complex explanations (if the additional dimensions of complexity also contain more predictive information). When designing \simevals{} and interpreting agent accuracy, one should factor in the complexity of the explanations that are being evaluated. If the explanation is complex and high-dimensional, then one might discount the agent's accuracy as a measure of informing human performance.  In our user study, we limited the dataset to 2 dimensions to control explanation complexity and reduce the likelihood that the user experiences cognitive overload. 

\textbf{User Interfaces.} The way information is presented to the user through the user study interface or visualization of the explanations can also affect a human's ability to make decisions~\citep{schnabel2018improving}. In our user study, we control these factors by designing a simple MTurk interface using Qualtrics (a widely-used survey platform).

\subsection{Future Work}

It is evident that ML models and humans learn and reason differently. While we found overall that the accuracy of an algorithmic agent can be interpreted as a measure of an explanation's utility for humans, we observe in our MTurk Study that there is a gap.  Our proposed framework does not intend to measure potential cognitive factors that may affect the ability of humans to use explanations.  An interesting and challenging direction for future work would attempt to (a) understand what such factors are, and (b) measure the extent to which a factor is present in the set of candidate explanations.  A better understanding of such factors could also inform development of new explanation methods that meet these desiderata. 

It's also worth noting that there are many domains/types of data where humans are still much better at learning than machines.  In these settings, an algorithmic agent would also fail to measure the utility of the explanation to a human. Future work may include bringing a human in the loop to define human strategies or behaviors that the agent should follow as to make the agent's reasoning more similar to a human's.





\end{document}